\documentclass{article}

    \PassOptionsToPackage{numbers, compress}{natbib}
    \usepackage[preprint]{neurips_2020}

\usepackage[utf8]{inputenc} % allow utf-8 input
\usepackage[T1]{fontenc}    % use 8-bit T1 fonts
\usepackage{hyperref}       % hyperlinks
\usepackage{url}            % simple URL typesetting
\usepackage{booktabs}       % professional-quality tables
\usepackage{amsfonts}       % blackboard math symbols
\usepackage{nicefrac}       % compact symbols for 1/2, etc.
\usepackage{microtype}      % microtypography

\usepackage{amsmath}
\usepackage{amsthm}
\usepackage[ruled, noend, noline]{algorithm2e}
\usepackage{mathtools}

\title{Distributed Sketching Methods for Privacy Preserving Regression} 

\author{%
  Burak Bartan \\
  Department of Electrical Engineering\\
  Stanford University\\
  Stanford, CA 94305 \\
  \texttt{bbartan@stanford.edu} \\
   \And
  Mert Pilanci \\
  Department of Electrical Engineering\\
  Stanford University\\
  Stanford, CA 94305 \\
  \texttt{pilanci@stanford.edu} \\
}

\theoremstyle{plain}

\newtheorem{theo}{Theorem}[section]

\newtheorem{lem}{Lemma}[section]
\newtheorem{prop}{Proposition}[section]
\newtheorem{cor}{Corollary}[section]

\theoremstyle{definition} 

\newtheorem{nota}{Notation}[section]
\newtheorem{de}{Definition}[section]
\newtheorem{exa}{Example}[section]
\newtheorem{as}{Assumption}[section]
\newtheorem{alg}{Algorithm}[section]

\newcommand{\btheo}{\begin{theo}}
\newcommand{\bde}{\begin{de}}
\newcommand{\ble}{\begin{lem}}
\newcommand{\bpr}{\begin{prop}}
\newcommand{\bno}{\begin{nota}}
\newcommand{\bex}{\begin{exa}}
\newcommand{\bcor}{\begin{cor}}
\newcommand{\spro}{\begin{proof}}
\newcommand{\bas}{\begin{as}}
\newcommand{\balg}{\begin{alg}}

\newcommand{\etheo}{\end{theo}}
\newcommand{\ede}{\end{de}}
\newcommand{\ele}{\end{lem}}
\newcommand{\epr}{\end{prop}}
\newcommand{\eno}{\end{nota}}
\newcommand{\eex}{\end{exa}}
\newcommand{\ecor}{\end{cor}}
\newcommand{\fpro}{\end{proof}}
\newcommand{\eas}{\end{as}}
\newcommand{\ealg}{\end{alg}}

\theoremstyle{plain}

\newtheorem{theos}{Theorem}
\newtheorem{props}{Proposition}
\newtheorem{lems}{Lemma}
\newtheorem{cors}{Corollary}

\theoremstyle{definition}
\newtheorem{exas}{Example}
\newtheorem{algs}{Algorithm}
\newtheorem{asss}{Assumption}
\newtheorem{defns}{Definition}

\newtheorem{rem}{Remark}

\newcommand{\btheos}{\begin{theos}}
\newcommand{\etheos}{\end{theos}}
\newcommand{\bprops}{\begin{props}}
\newcommand{\eprops}{\end{props}}
\newcommand{\bdes}{\begin{defns}}
\newcommand{\edes}{\end{defns}}
\newcommand{\blems}{\begin{lems}}
\newcommand{\elems}{\end{lems}}
\newcommand{\bcors}{\begin{cors}}
\newcommand{\ecors}{\end{cors}}
\newcommand{\bexs}{\begin{exas}}
\newcommand{\eexs}{\end{exas}}
\newcommand{\balgs}{\begin{algs}}
\newcommand{\ealgs}{\end{algs}}
\newcommand{\bass}{\begin{asss}}
\newcommand{\eass}{\end{asss}}

\newcommand{\bremark}{\begin{rem}}
\newcommand{\eremark}{\end{rem}}

\newcommand{\lambdamax}[1]{\ensuremath{\lambda_{max}}}

\newcommand{\xstar}{\ensuremath{x^*}}

\newcommand{\mprob}{\ensuremath{\mathbb{P}}}

\newcommand{\real}{\ensuremath{\mathbb{R}}}
\newcommand{\defn}{\ensuremath{: \, =}}

\newcommand{\Exs}{\ensuremath{\mathbb{E}}}

\DeclareMathOperator{\tr}{tr} %%%%

\begin{document}

\maketitle

\begin{abstract}
   In this work, we study distributed sketching methods for large scale regression problems. We leverage multiple randomized sketches for reducing the problem dimensions as well as preserving privacy and improving straggler resilience in asynchronous distributed systems.  We derive novel approximation guarantees for classical sketching methods and analyze the accuracy of parameter averaging for distributed sketches. We consider random matrices including Gaussian, randomized Hadamard, uniform sampling and leverage score sampling in the distributed setting. Moreover, we propose a hybrid approach combining sampling and fast random projections for better computational efficiency. We illustrate the performance of distributed sketches in a serverless computing platform with large scale experiments.
\end{abstract}
\section{Introduction}
We investigate distributed sketching methods in large scale regression problems. In particular, we study parameter averaging for variance reduction and establish theoretical results on approximation performance. %In particular, we discuss the application of parameter averaging to linear least squares, iterative Hessian sketch, and Lasso. 
Employing parameter averaging in distributed systems enables asynchronous updates, since a running average of available parameters can approximate the result without waiting for all workers to finish their jobs. Moreover, sketching provably preserves the privacy of the data, making it an attractive choice for massively parallel cloud computing.

We consider both underdetermined and overdetermined linear regression problems in the regime where the data does not fit in main memory. Such linear regression problems and linear systems are commonly encountered in a multitude of problems ranging from statistics and machine learning to optimization. Being able to solve large scale linear regression problems efficiently is crucial for many applications. In this paper, the setting we consider is a distributed system with $q$ workers that run in parallel. Applications of randomized sketches and dimension reduction to linear regression and other optimization problems has been extensively studied in the recent literature by works including \cite{DriMahMutSar09}, \cite{mahoney2018averaging}, \cite{PilWai14b}, \cite{lacotte2019faster}, \cite{Mahoney11}, \cite{PilWai14a}. In this work, we investigate averaging the solutions of sketched sub-problems. This setting for overdetermined problems was also studied by \cite{mahoney2018averaging}. In addition, we also consider regression problems where the number of data samples is less than the dimensionality and investigate the properties of averaging such problems.

An important advantage of randomized sketching in distributed computing is the independent and identically distributed nature of all of the computational tasks. Therefore, sketching offers a resilient computing model where node failures, straggler workers as well as additions of new nodes can be easily handled, e.g., generating more data sketches. An alternative to averaging that would offer similar benefits is the asynchronous SGD algorithm \cite{bertsekas1989,recht2011hogwild}. However, the convergence rates of asynchronous SGD methods necessarily depend on the properties of the input data such as its condition number \cite{bertsekas1989,recht2011hogwild}. In contrast, as we show in this work, distributed sketching and averaging has stronger convergence guarantees that do not depend on the condition number of the data matrix.

In order to illustrate straggler resilience, we have implemented distributed sketching for AWS Lambda, which is a serverless computing platform. Serverless computing is an emerging architectural paradigm that presents a compelling option for parallel data intensive problems. One can access to a high number of serverless compute functions each with limited resources relatively inexpensively. If the computation task can be factored into many sub-problems with no communication in-between, serverless computing can offer many advantages over server-based computing due to its scalability and pricing model. Given that in this work we are interested in solving large scale problems with number of data samples much higher than the dimensionality, it suits to use serverless computing in solving these types of problems using model averaging. We discuss how distributed sketching methods scale and perform for different large scale datasets on AWS Lambda \cite{jonas2017pywren}. 

Data privacy is an increasingly important issue in cloud computing, one that has been studied in recent works including \cite{zhou2009privacy}, \cite{showkatbakhsh2018privacy}, \cite{ZhoLafWas07}, \cite{blocki12jltprivacy}. Distributed sketching comes with the benefit of privacy preservation. To clarify, let us consider a setting where the master node computes sketched data $S_kA$, $S_kb$, $k=1,...,q$ locally where $S_k \in \mathbb{R}^{m\times n}$ are the sketching matrices and $A \in \mathbb{R}^{n \times d}$, $b \in \mathbb{R}^{n}$ are the data matrix and the output vector, respectively for the regression problem $\min_x\|Ax-b\|_2^2$. The master node sends only the sketched data to worker nodes for computational efficiency, as well as preserving data privacy. In particular, the mutual information as well as differential privacy can be controlled when we reveal $S_kA$ and keep $A$ hidden. Furthermore, one can trade privacy for accuracy by choosing a suitable sketch dimension $m$. 
%\todo[inline]{mutual info bound here -- should I remove the bound from section III.A (privacy.tex) ?}
% \begin{align}
%     \frac{I(S_kA; A)}{nd} \leq \frac{m}{n} \log (2\pi e \gamma^2).
% \end{align}

We study various sketching methods in the context of averaging including Gaussian sketch, randomized orthonormal systems based sketch, uniform sampling, and leverage score sampling. In addition, we discuss a hybrid sketching approach and illustrate its performance through numerical examples.

\subsection{Related Work and Main Contributions}
The work \cite{mahoney2018averaging} investigates model averaging for regression from optimization and statistical perspectives. Our work studies sketched model averaging from the optimization perspective. The most relevant result of \cite{mahoney2018averaging}, using our notation, can be stated as follows. Setting the sketch dimension $m=\mathcal{O}(\mu d (\log d)/ \epsilon )$ for uniform sampling (where $\mu$ is row coherence) and $m=\tilde{\mathcal{O}}(d/\epsilon )$ (with $\tilde{\mathcal{O}}$ concealing logarithmic factors) for other sketches, the inequality $f(\bar{x}) - f(x^*) \leq (\epsilon/q+\epsilon^2 ) f(x^*)$ holds with high probability. According to this result, for large $q$, the cost of the averaged solution $f(\bar{x})$ will be less than $\epsilon^2 f(x^*)$. We prove in this work that for Gaussian sketch, $\Exs[f(\bar{x})]$ will converge to the optimal cost $f(x^*)$ as $q$ increases (i.e. unbiasedness). We also identify the exact expected error for a given number of workers $q$ for Gaussian sketch. 

We show that the expected difference between the costs for the averaged solution and the optimal solution has two components, namely variance and bias squared (see Lemma \ref{expected_obj_val_diff}). This result implies that for Gaussian sketch, which we prove to be unbiased (see Lemma \ref{gaussian_one_sketch}), the number of workers required for a given error $\epsilon$ scales with $1/\epsilon$. For the Hogwild algorithm \cite{recht2011hogwild}, which is also asynchronous, but addresses a more general class of problems, the number of iterations required for error $\epsilon$ scales with $\log(1/\epsilon)/\epsilon$ and also depends on the input data.

%Our analysis approach allows the cost to be decomposed into bias and variance. 
We derive upper bounds for the estimator biases for additional sketching matrices including randomized orthonormal systems (ROS) based sketch, uniform sampling, and leverage score sampling sketches. Analysis of the bias term is critical in understanding how close to the optimal solution we can hope to get and how bias depends on the sketch dimension $m$.

\vspace{-2mm}
\section{Preliminaries}
We consider the linear least squares problem given by $\xstar = \arg\min_{x}f(x)$ where $f(x)=\|Ax-b\|^2_2$,
% \begin{align}
% \xstar = \arg\min_{x} \underbrace{ \|Ax-b\|^2_2}_{f(x)}\,,
% \end{align}
$A\in \real^{n\times d}$ is the data matrix and $b \in \real^{n}$ is the output vector. We assume a distributed setting with $q$ workers running in parallel and a central node (i.e. master node) that forms the averaged solution $\bar x$:
\begin{align}
\bar x \defn \frac{1}{q} \sum_{k=1}^q \hat x_k, \text{ where } \hat x_k = \arg\min_{x} \|S_kAx-S_kb\|^2_2 \,,
\end{align}
where $\hat{x}_k$ is the output of the $k$'th worker.
% \begin{align*}
% \hat x_k = \arg\min_{x} \|S_k(Ax-b)\|^2_2\,.
% \end{align*}
The matrices $S_k\in \real^{m\times n}$ are independent random sketch matrices that satisfy the scaling $\Exs [S_k^TS_k] = I_n$. The algorithm is listed in Algorithm \ref{distrib_avg_alg}.

% zero ya set etmeyi kaldiriyorum
% Let us define the following event for a single sketch: $E=\{(1+\epsilon)A^TA \succeq A^TS^TSA \succeq (1-\epsilon)A^TA\}$. Let $\tilde{x}$ denote the solution due to a sketch:
% \begin{align*}
%     \tilde{x} = (A^TS^TSA)^{-1}A^TS^TSy\mathbb{I}_E + 0_d\mathbb{I}_{E^C},
% \end{align*}
% where $0_d$ stands for the all zeros vector of length $d$. 

We will use the following equivalence of expression in the sequel: The expected difference between the costs for the averaged solution $f(\bar{x})$ and the optimal solution $f(x^*)$ is equal to:
\begin{align} \label{eq:expected_diff}
    \Exs[f(\bar{x})]-f(x^*)% &= \Exs[\|A\bar{x}-b\|_2^2]-f(x^*) \\
    &= \Exs[\|A(\bar{x}-x^*)+Ax^*-b\|_2^2]-f(x^*) \nonumber \\
    & = \Exs [\|A(\bar{x}-x^*)\|_2^2 + \|Ax^*-b\|_2^2]-f(x^*) \nonumber \\
    & = \Exs [\|A(\bar{x}-x^*)\|_2^2]\,,
\end{align}
where we have used the orthogonality property of the optimal least squares solution $x^*$ given by the normal equations $A^T (Ax^*-b) = 0$.%, which implies $(\bar x - x^*)^T A^T (Ax^*-b) = 0$.

Throughout the text, the notation we use for the SVD of $A \in \mathbb{R}^{n \times d}$ is $A=U\Sigma V^T$. Whenever $n \geq d$, we assume $A$ has full column rank, hence the dimensions for $U,\Sigma,V$ are as follows: $U \in \mathbb{R}^{n \times d}$, $\Sigma \in \mathbb{R}^{d \times d}$, and $V^T \in \mathbb{R}^{d \times d}$. We use the row vector $\tilde{u}_i^T \in \mathbb{R}^{1\times d}$ to refer to the $i$'th row of $U$. We use $S$ and $S_k$ to denote sketching matrices.

We give the proofs of all the lemmas and theorems in the appendix. 

\vspace{-4mm}
\DontPrintSemicolon
\begin{algorithm}
 \KwIn{Data matrix $A \in \mathbb{R}^{n\times d}$, target vector $b \in \mathbb{R}^{n}$.}
 \textbf{Workers $k=1,...,q$ in parallel: } \;
 %Sample $S_k \in \mathbb{R}^{m\times n}$. \;
 Obtain the sketched data and sketched output: $S_kA$ and $S_kb$. \;
 Solve the problem $\hat x_k = \arg\min_{x} \|S_kAx-S_kb\|^2_2$, and send $\hat x_k$ to the master node. \;
 \textbf{Master:} return $\bar x = \frac{1}{q} \sum_{k=1}^q \hat x_k$. \;
 \caption{Distributed sketching algorithm.}
 \label{distrib_avg_alg}
\end{algorithm}
\vspace{-4mm} 
\section{Gaussian Sketch}
In this section, we present results for the error of Algorithm \ref{distrib_avg_alg} when the sketch matrix is Gaussian, i.e., the entries of $S_k$ are distributed as i.i.d. Gaussian. We first obtain a characterization of the expected prediction error of a single sketched least squares solution. 
\blems \label{gaussian_one_sketch}
For the Gaussian sketch with $m>d+1$, the estimator $\hat{x}_k$ satisfies
\begin{align*}
    \Exs [\| A (\hat{x}_k - x^*) \|_2^2] = \Exs[ f(\hat{x}_k) ] - f(x^*) =  f(x^*) \frac{d}{m-d-1}.
\end{align*}
\elems
To the best of our knowledge, this result is novel in the theory of sketching. Existing results (see e.g. \cite{Mahoney11,mahoney2018averaging,woodruff2014sketching}) characterize a high probability upper bound on the error, whereas the above is a sharp and exact formula for the expected squared norm of the error. Theorem \ref{Thm:AvgGaussian} builds on Lemma \ref{gaussian_one_sketch} to establish the error for the averaged solution.

% \btheos[Cost Approximation]
% Let each sketch dimension be chosen as $m=\frac{2rank(A)}{q\epsilon}$ using Gaussian sketches then the parallel scheme runs in time $O(\frac{rank(A)^2 d}{q \epsilon })$ and gives an approximation
% \begin{align*}
% \frac{\Exs f(\bar x) - f(\xstar)}{f(\xstar)}\le \epsilon 
% \end{align*}
% where $0<\epsilon\le \frac{1}{2q}$ is arbitrary.
% \etheos

% \section{Distributed Gaussian Sketch}
% We now leverage the exact expression in the previous section to obtain a sharp and exact result for averaging multiple sketches.
% \btheos[Cost Approximation]
% \label{Thm:AvgGaussian}
% Let each sketch dimension be chosen as $m=\frac{2rank(A)}{q\epsilon}$ using Gaussian sketches then the parallel scheme runs in time $O(\frac{rank(A)^2 d}{q \epsilon })$ and \begin{align*}
%     \frac{\Exs[f(\bar{x})]-f(x^*)}{f(x^*)} = \frac{1}{q} \frac{d}{m-d-1} 
% \end{align*}
% given that $A^TS_k^TS_kA \succ 0$ for all $k=1,...,q$. Equivalently, 
% \begin{align*}
%     \frac{f(\bar{x})-f(x^*)}{f(x^*)} \leq  \frac{\epsilon}{q}
% \end{align*}
% holds with probability at least $(1-e^{-m})^q \left(1 - \frac{1}{\epsilon} \frac{d}{m-d-1} \right)$.
% \etheos

\btheos[Cost Approximation]
\label{Thm:AvgGaussian}
Let $S_k$, $k=1,...,q$ be Gaussian sketches, then Alg. \ref{distrib_avg_alg} runs in time $O(md^2)$, and the error of the averaged solution $\bar{x}$ satisfies
% Let each sketch dimension be chosen as $m=\frac{2rank(A)}{q\epsilon}$ using Gaussian sketches then the parallel scheme runs in time $O(\frac{rank(A)^2 d}{q \epsilon })$ and
\begin{align*}
    \frac{\Exs[f(\bar{x})]-f(x^*)}{f(x^*)} = \frac{1}{q} \frac{d}{m-d-1} 
\end{align*}
given that $A^TS_k^TS_kA \succ 0$ for all $k=1,...,q$. Equivalently, $
    \frac{f(\bar{x})-f(x^*)}{f(x^*)} \leq  \frac{\epsilon}{q} $
holds with probability at least $(1-e^{-c_1m})^q \left(1 - \frac{1}{\epsilon} \frac{d}{m-d-1} \right)$.
\etheos

Proofs of Lemma \ref{gaussian_one_sketch} and Theorem \ref{Thm:AvgGaussian} mainly follow from the results on inverse-Wishart distribution, and the proofs are given in the Appendix. Theorem \ref{Thm:AvgGaussian} shows that the expected error scales with $1/q$, and as $q$ increases the error goes to $0$ (i.e. unbiased). The unbiasedness property is not observed for other sketches such as uniform sampling, which is discussed in Section \ref{sec:other_sketching_matrices}.

\subsection{Privacy Preserving Property}
We now digress from the convergence properties of Gaussian sketch to state the results on privacy. We use the notion of differential privacy given in Definition \ref{def:diff_privacy} \cite{dwork06privacy}. Differential privacy is a worst case type of privacy notion which does not require distribution assumptions on the data. It has been the privacy framework adopted by many works in the literature. %Theorem \ref{thm:diff_priv_cited} from \cite{sheffet15privacy} indicates that publishing the product of a data matrix and a Gaussian random projection matrix preserves $(\varepsilon, \delta)$-differential privacy if the minimum singular value of the data matrix is at least $w$, which is defined in \eqref{eq:def_w}.

\bdes[$(\varepsilon, \delta)$-Differential Privacy, \cite{dwork06privacy}, \cite{sheffet15privacy}] \label{def:diff_privacy}
An algorithm ALG which maps $(n\times d)$-matrices into some range
$\mathcal{R}$ satisfies $(\varepsilon, \delta)$-differential privacy if for all pairs of neighboring inputs $A$ and $A^\prime$ (i.e. they differ only in a single row) and all subsets $\mathcal{S} \subset \mathcal{R}$, it holds that $P(\text{ALG}(A) \in \mathcal{S}) \leq e^\varepsilon P(\text{ALG}(A^\prime) \in \mathcal{S}) + \delta$. %When $\delta=0$, we say the algorithm is $\varepsilon$-differentially private.
\edes

%\todo[inline]{Bunu theorem olarak yazalim ama $mq \le O(\frac{constant}{\varepsilon^2})$ gibi simplified formda. buradaki derivasyon supp. file'da olsun.}

%\todo[inline]{Bu privacy right sketchte nasil oluyor orada da yazabiliriz.}

%\todo[inline]{Main paperdaki butun prooflari supp. file'a koymak en iyisi gibi. ornegin theorem 1, lemma 1, lemma 7.}

% \todo[inline]{Bu theoremi su sekilde yazabiliriz. For $\varepsilon,\delta$ DP with respect to all workers, pick $m = \frac{1}{q} \frac{1}{24} \left(\left(\frac{\sigma_{\min}^2}{B^2} - 1\right) \frac{3\varepsilon}{3+\varepsilon} - 6 \right)^2 $. For $\varepsilon,\delta$ DP with respect to a single worker pick... (1/q olmayan hali). Butun detaylari ve karisik denklemleri supp. file'a koyalim. Yukaridaki 1/$\varepsilon^2$ approximation ratio'yu remark olarak yazalim ve konustugumuz referansla karsilastiralim. Orada daha kotu bir dependence olmasi lazim.}

Theorem \ref{thm:diff_priv_result} characterizes the required conditions and the sketch size to guarantee $(\varepsilon, \delta)$-differential privacy for a given dataset. The proof of Theorem \ref{thm:diff_priv_result} is based on the $(\varepsilon, \delta)$-differential privacy result for i.i.d. Gaussian random projections from \cite{sheffet15privacy}.
%We note that the values $m$ given in Theorem \ref{thm:diff_priv_result} are upper bounds for $m$, and ...

\btheos[] \label{thm:diff_priv_result}
Given the data matrix $A \in \mathbb{R}^{n\times d}$ and the output vector $y \in \mathbb{R}^{n}$, let $A_c$ denote the concatenated version of $A$ and $b$, i.e., $A_c = [A, b] \in \mathbb{R}^{n\times {(d+1)}}$. Set $\delta = 4/e^\beta$. Let $B=B_0 \sqrt{d}$ and $\sigma_{\min}(A_c)=\sigma_0 \sqrt{n}$, where $B_0$ and $\sigma_0$ satisfy $|A_{c,ij}| \leq B_0$ for all $i,j$, and $\sigma_{\min}(\frac{1}{n}A_c^TA_c) = \sigma_0^2$. If the condition
\begin{align} \label{eq:diff_priv_condition}
    \frac{n}{d} \geq \left( 3 + \frac{2\beta}{\varepsilon} \right) \frac{B_0^2}{\sigma_0^2}
    %\sigma_{\min} \geq B \sqrt{3 + \frac{2\beta}{\varepsilon}}
\end{align}
is satisfied, then we pick the sketch size as
\begin{align} \label{eq:sketch_size_expressions}
    m &= O\left(\beta \frac{n^2}{d^2} \frac{\varepsilon^2}{(\varepsilon + \beta)^2}\right) \quad\text{for privacy w.r.t. a single worker,} \\
    %\left\lfloor \frac{\beta}{8} \left(\left(\frac{\sigma_{\min}^2}{B^2} - 1\right) \frac{\varepsilon }{\varepsilon + \beta} - 2 \right)^2 \right\rfloor \quad\text{for privacy w.r.t. a single worker,} \\
    m &= O\left(\frac{\beta}{q} \frac{n^2}{d^2} \frac{\varepsilon^2}{(\varepsilon + \beta)^2}\right) \quad\text{for privacy w.r.t. $q$ workers,}
    %\left\lfloor \frac{1}{q}\frac{\beta}{8} \left(\left(\frac{\sigma_{\min}^2}{B^2} - 1\right) \frac{\varepsilon }{\varepsilon + \beta} - 2 \right)^2 \right\rfloor \quad\text{for privacy w.r.t. $q$ workers},
\end{align}
and then publishing the sketched matrices $S_kA_c \in \mathbb{R}^{m\times {(d+1)}}$, $k=1,...,q$, provided that $d+1 < m$, is $(\varepsilon, \delta)$-differentially private, where $S_k \in \mathbb{R}^{m\times (n+1)}$ is the Gaussian sketch, $\varepsilon > 0$, $\beta > 1+\ln(4)$.
\etheos

%According to Theorem \ref{thm:diff_priv_result}, provided that the conditions are satisfied, we can pick the sketch size values as given in \eqref{eq:sketch_size_expressions} for $(\varepsilon, \delta)$-differential privacy. 
% Burayi cikarabiliriz
%In the cases where one is not able to pick $m$ that would satisfy the privacy requirements, the authors in \cite{sheffet15privacy} utilize the method of regularization as a way of ensuring privacy, which corresponds to appending a scaled identity matrix to the data matrix. The scaling factor of the identity is chosen to satisfy the differential privacy requirements.

% \bremark
% Theorem \ref{thm:diff_priv_result} in \eqref{eq:diff_priv_condition} implies that if $\frac{n}{d}$ is on the order of $O(\frac{\beta}{\varepsilon})$, then we can pick $m = O\left(\frac{\beta}{q} \frac{n^2}{d^2} \frac{\varepsilon^2}{(\varepsilon + \beta)^2}\right)$ for $(\varepsilon, \delta)$-differential privacy with respect to all $q$ workers.
% \eremark

\bremark
For fixed values of $\beta$, $\sigma_{\min}$, $B$, and $d$, the approximation error is on the order of $O\left(\frac{1}{\varepsilon^2} \right)$ for $(\varepsilon, \delta)$-differential privacy with respect to all $q$ workers.
\eremark

The authors in \cite{huang15distprivacy} show that under $\varepsilon$-differential privacy (i.e. equivalent to Definition \ref{def:diff_privacy} with $\delta=0$), the approximation error of their distributed iterative algorithm is on the order of $O(\frac{1}{\varepsilon^2})$, which is on the same order as Algorithm \ref{distrib_avg_alg}. The two algorithms however have different dependencies on parameters, which are hidden in the $O$-notation. %For instance, the error of the algorithm in \cite{huang15distprivacy} has dependence on the order of $O(\sigma_{\max}^4)$ while Algorithm \ref{distrib_avg_alg} has dependence on the order of $O(B^4 / \sigma_{\min}^4)$. 
The approximation error of the algorithm in \cite{huang15distprivacy} depends on parameters that Algorithm \ref{distrib_avg_alg} does not have such as the initial step size, the step size decay rate, and noise decay rate. The reason for this is that the algorithm in \cite{huang15distprivacy} is a synchronous iterative algorithm designed to solve a more general class of optimization problems. Algorithm \ref{distrib_avg_alg}, on the other hand, is designed to solve linear regression problems and this makes it possible to design Algorithm \ref{distrib_avg_alg} as a single-round communication algorithm. This also establishes Algorithm \ref{distrib_avg_alg} as more robust against slower workers with higher straggler-resilience.

\bremark
For the sketching method for least-norm problems (i.e., right-sketch) discussed in Section \ref{sec:right_sketch}, a similar privacy statement holds. In right sketch, we only sketch the data matrix $A$ and not the output vector $b$. For publishing $AS_k^T$ to be $(\varepsilon, \delta)$-differentially private, Theorem \ref{thm:diff_priv_result} still holds with the modification that we replace $A_c$ with $A^T$.
\eremark

% S is scaled R
% used \varepsilon in two different places
% consider all workers ++
% A should be: concatenated [A, y]: n x (d+1) ++
% m is an integer ++

%%%% papers on differential privacy:
% https://ieeexplore.ieee.org/stamp/stamp.jsp?arnumber=8437722
% https://arxiv.org/pdf/0901.1365.pdf
% https://arxiv.org/pdf/1204.2136.pdf
% https://arxiv.org/pdf/1507.02482.pdf
% Private Approximations of the 2nd-Moment Matrix Using Existing Techniques in Linear Regression: https://arxiv.org/abs/1507.00056 

%%%% papers 06/02 %%%%
% Differentially Private Distributed Optimization
% Randomness Efficient Fast-Johnson-Lindenstrauss Transform with Applications in Differential Privacy
% A Private and Finite-Time Algorithm for Solving a Distributed System of Linear Equations
% Privacy-Preserving Distributed Optimization via Subspace Perturbation: A General Framework
% Acceleration and Privacy Protection for Distributed Computing

\section{Other Sketching Matrices}
\label{sec:other_sketching_matrices}
In this section, we consider three additional sketching matrices, namely, randomized orthonormal systems (ROS) based sketching, uniform sampling, and leverage score sampling. For each of these sketches, we present upper bounds on the norm of the bias. In particular, the results of this section focus on the bias bounds for a single output $\hat{x}_k$, and the way these results are related to the averaged solution $\bar{x}$ is through Lemma \ref{expected_obj_val_diff}. Lemma \ref{expected_obj_val_diff} expresses the expected objective value difference in terms of bias and variance of a single estimator. It is possible to obtain high probability bounds for $(f(\bar{x})-f(x^*))/f(x^*)$ for these other sketches based on the bias bounds given in this section, using an argument similar to the one given in the proof of Theorem \ref{Thm:AvgGaussian}. This approach would involve defining an event that bounds the error for the single sketch estimator. %We skip the details in this version of the work.

% \blems \label{expected_obj_val_diff}
% For any i.i.d. sketching matrices $S_k$, $k=1,...,q$, the expected objective value difference is upper bounded by
% \begin{align} \label{approx_guarantee}
%     &\Exs[f(\bar{x})]-f(x^*)  \nonumber \\ &\leq \frac{4}{q}\Exs[\|A\hat{x}_1 - Ax^* \|_2^2]  +  (2+\frac{4}{q}) \|\Exs[A\hat{x}_1]-Ax^*\|_2^2,
% \end{align}
% where $\hat{x}_1$ is the solution returned by worker $k=1$ (in fact it could be any of the worker outputs $\hat{x}_k$ as they are all statistically identical).
% \elems
% \begin{proof}
% Proof at the appendix.
% \end{proof}

% We prove that the bias term $\|\Exs[A\hat{x}_1]-Ax^*\|_2^2$ is zero in the case of Gaussian sketch and derive upper bounds for the bias for other sketches. We first derive results for the Gaussian sketch and then move on to analyzing other sketches which include randomized orthogonal systems based sketch, uniform sampling sketch and leverage score sampling sketch.

\blems \label{expected_obj_val_diff}
For any i.i.d. sketching matrices $S_k$, $k=1,...,q$, the expected objective value difference can be decomposed as
\begin{align} \label{approx_guarantee}
    \Exs[f(\bar{x})]-f(x^*) = \frac{1}{q} \Exs \left[ \|A\hat{x}_1-Ax^* \|_2^2 \right] + \frac{q-1}{q} \| \Exs[A\hat{x}_1] - Ax^* \|_2^2,
\end{align}
where $\hat{x}_1$ is the solution returned by worker $k=1$ (in fact it could be any of the worker outputs $\hat{x}_k$ as they are all statistically identical).
\elems

Lemma \ref{norm_of_bias} establishes an upper bound on the norm of the bias for any i.i.d. sketching matrix.
\blems \label{norm_of_bias}
Let the SVD of $A$ be denoted as $A = U\Sigma V^T$. Let $z \defn U^TS_k^TS_kb^\perp$ and $Q \defn (U^TS_k^TS_kU)^{-1} $ where $b^\perp \defn b - Ax^*$. For $S_k$ any sketch matrix with $\Exs[S_k^TS_k]=I_n$, under the assumption that $(1-\epsilon)I_d \preceq Q \preceq (1+\epsilon)I_d$, the norm of the bias for a single sketch is upper bounded as:
\begin{align}
    \|\Exs[A\hat{x}_k] - Ax^*\|_2 \leq \sqrt{4\epsilon \Exs[\| z\|_2^2]}.
\end{align}
\elems

We note that Lemma \ref{expected_obj_val_diff} and \ref{norm_of_bias} apply to all of the sketching matrices considered in this work. We now move on to state upper bound results for the bias of various sketching matrices. In terms of the computational complexity required to sketch data, uniform sampling is the most inexpensive one and it is also the one with the highest bias as we will see in the following subsections.

% \begin{proof}
% Proof at the appendix.
% \end{proof}

% We note that the second term in the bound in Lemma \ref{expected_obj_val_diff} is the norm of the bias for which we have a bound (Lemma \ref{norm_of_bias}). We need to come up with a bound for the first term as well.

% For properly chosen sketch dimension $m$, we have the following approximation guarantee with  probability at least $1-c_1e^{-c_2m\epsilon^2}$ [cite this result]:
% \begin{align} \label{approx_eq}
%     f(\hat{x}_1) &\leq (1+\epsilon)^2 f(x^*) \\
%     f(\hat{x}_1) - f(x^*) &\leq (2\epsilon + \epsilon^2) f(x^*) \nonumber \\ 
%     f(\hat{x}_1) - f(x^*) &\leq 3\epsilon f(x^*) \nonumber \\
%     f(\hat{x}_1) - f(x^*) &\leq \epsilon^\prime f(x^*) \nonumber
% \end{align}

% We rewrite the first term in  \eqref{approx_guarantee} as
% \begin{align}\label{cost_approx}
%     &\Exs[||A\hat{x}_1-Ax^*||_2^2] = \Exs[f(\hat{x}_1) - f(x^*)] \nonumber \\
%     &\leq \epsilon^\prime f(x^*) (1-c_1 e^{-c_2m{\epsilon^\prime}^2}) + ||Ax^*||_2^2 c_1 e^{-c_2m{\epsilon^\prime}^2} 
% \end{align}
% If the guarantee in \eqref{approx_eq} is not satisfied we know the event $E$ does not take place, hence the estimator $\hat{x}_1$ is equal to the zero vector. This is the reasoning behind the bound given in the second term of \eqref{cost_approx}.

% Can we say if $f(\hat{x}_1) - f(x^*) \leq \delta^\prime f(x^*)$ is not satisfied, then we set the estimator $\hat{x}_1$ to the zero vector so we can have the bound $||A(\hat{x}_1-Ax^*)||_2^2 \leq ||Ax^*||_2^2$? -- yes

\subsection{Randomized Orthonormal Systems (ROS) based Sketches}
ROS-based sketches can be represented as $S=PHD$ where $P$ is for sampling $m$ rows out of $n$, $H$ is the ($n\times n$)-dimensional Hadamard matrix, and $D$ is a diagonal matrix with diagonal entries sampled from the Rademacher distribution. This type of sketching matrix is also known as randomized Hadamard based sketch and it has the advantage of lower computational complexity ($O(n\log n)$) over Gaussian sketch ($O(n^3)$). %The scaling of $S$ is such that $\Exs [S^T S]=\frac{1}{m} \sum_{i=1}^m \Exs[s_is_i^T] = I_n$.
\blems \label{bound_z_norm_ROS}
Let $S_k$ be the ROS sketch and $z= U^T S_k^T S_k b^\perp$. Then we have
\begin{align}
    \Exs [\|z\|_2^2] \leq \frac{d}{m} \left(1-\frac{2\min_i \|\tilde{u}_i \|_2^2}{d} \right) f(x^*), \quad \text{and} \quad
    \|\Exs[A\hat{x}_k] - Ax^*\|_2 \leq \sqrt{4\epsilon \frac{d}{m} f(x^*)}.
\end{align}
\elems

\subsection{Uniform Sampling}
%The sketching matrix $S$ for uniform sampling is equal to $P$. 
In uniform sampling, each row of $S$ consists of a single $1$ and $(n-1)$ $0$'s and the position of $1$ in every row is distributed according to uniform distribution. Then, the $1$'s in $S$ are scaled so that $\Exs[S^TS]=I_n$. %Note that we scale $S$ such that $\Exs [S^T S]=\frac{1}{m} \sum_{i=1}^m \Exs[s_is_i^T] = I_n$. 
%In fact for $\Exs[s_is_i^T] = I_n$, and $s_i$'s are from the set of standard basis vectors scaled by $\sqrt{n}$.
We note that the bias for uniform sampling is different when it is done with or without replacement. Lemma \ref{bound_z_norm_uniform} gives bounds for both cases and verifies that the bias of $\hat{x}_k$ is less if sampling is done without replacement.

\blems \label{bound_z_norm_uniform}
Let $S_k$ be the sketching matrix for uniform sampling and $z= U^T S_k^T S_k b^\perp$. Then we have
\begin{align}
    \Exs [\|z\|_2^2] &\le \frac{n}{m}\, f(x^*) \max_i \|\tilde{u}_i \|_2^2, \quad \text{and} \quad \|\Exs[A\hat{x}_k] - Ax^*\|_2 \leq \sqrt{4\epsilon \frac{n}{m}\, f(x^*) \max_i \|\tilde{u}_i \|_2^2} \\
    \Exs[\|z\|_2^2] &\leq \frac{n}{m} \frac{n-m}{n-1}\, f(x^*) \max_i \|\tilde{u}_i \|_2^2,  \text{ and }  \|\Exs[A\hat{x}_k] - Ax^*\|_2 \leq \sqrt{4\epsilon \frac{n}{m} \frac{n-m}{n-1}\, f(x^*) \max_i \|\tilde{u}_i \|_2^2},
\end{align}
for sampling with and without replacement, respectively. %The norm of the bias satisfies
% \begin{align}
%     \|\Exs[A\hat{x}_k] - Ax^*\|_2 &\leq \sqrt{4\epsilon \frac{n}{m}\, f(x^*) \max_i \|\tilde{u}_i \|_2^2},  \\
%     \|\Exs[A\hat{x}_k] - Ax^*\|_2 &\leq \sqrt{4\epsilon \frac{n}{m} \frac{n-m}{n-1}\, f(x^*) \max_i \|\tilde{u}_i \|_2^2} ,
% \end{align}
% for sampling with and without replacement, respectively. 
\elems
% It follows from Lemma \ref{bound_z_norm_uniform} and Lemma \ref{norm_of_bias} that for uniform sampling based sketches, the norm of the bias is upper bounded by
% \begin{align*}
%     \|\Exs[A\hat{x}_k] - Ax^*\|_2 &\leq \sqrt{4\epsilon \frac{n}{m}\, f(x^*) \max_i \|\tilde{u}_i \|_2^2}.
% \end{align*}
\subsection{Leverage Score Sampling}
Row leverage scores of a matrix $A=U\Sigma V^T$ are given by $\ell_i = \|\tilde{u}_i\|_2^2$ for $i=1,...,n$ where $\tilde{u}_i$ denotes the $i$'th row of $U$. There is only one nonzero element in every row of the sketching matrix $S$ and the probability that the $j$'th entry of $s_i$ is nonzero is proportional to the leverage score $\|\tilde{u}_j\|_2^2$. That is, $\mprob[s_{ij}\neq 0] = \frac{\|\tilde{u}_j\|_2^2}{\sum_{j=1}^n \|\tilde{u}_j\|_2^2}$. The denominator is equal to $d$ because it is equal to the Frobenius norm of $U$ and the columns of $U$ are normalized. So, $\mprob[s_{ij}\neq 0] = \frac{1}{d}\|\tilde{u}_j\|_2^2=\frac{1}{d}\ell_i$. %We again scale the entries such that $\Exs [S^T S]=\frac{1}{m} \sum_{i=1}^m \Exs[s_is_i^T] = I_n$. %Hence, the nonzero entries ware set to $\sqrt{\frac{d}{\ell_i}}$.

\blems \label{bound_z_norm_leverage}
Let $S_k$ be the leverage score sampling based sketch and $z= U^T S_k^T S_k b^\perp$. Then we have
\begin{align}
    \Exs [\|z\|_2^2]  \leq \frac{d}{m} f(x^*), \quad \text{and}\quad \|\Exs[A\hat{x}_k] - Ax^*\|_2 \leq \sqrt{4\epsilon \frac{d}{m} f(x^*)}.
\end{align}
\elems

% Using Lemma \ref{bound_z_norm_leverage} and Lemma \ref{norm_of_bias}, for leverage score sampling based sketches, we obtain
% \begin{align*}
%     \|\Exs[A\hat{x}_k] - Ax^*\|_2 \leq \sqrt{4\epsilon \frac{d}{m} f(x^*)}.
% \end{align*}

\subsection{Hybrid Sketch}
% We start by motivating the idea of hybrid sketch. Let us assume that a worker in a distributed computing system cannot solve least squares problems for $A$ with number of rows more than some $n_{ls}$ for fixed $d$ due to limited memory. We then fix the sketching dimension to $m=n_{ls}$. Assuming that the entire dataset $A=[A_1^T,A_2^T,...,A_q^T]^T$ cannot be loaded into the memory, one way to sketch the entire dataset would be to load it one part $A_i$ at a time and apply sketching to that part (say SJLT). Depending on the size of $A$, this might take too long, so an alternative would be to read only randomly selected $n_{ls}$ rows of the data in the memory. This would be much faster since we read only $m$ rows of data, the same number of rows to be used in the least squares problem.

% These two scenarios in fact constitute the two extreme ends of what we could do in terms of the amount of data we read into the memory. Hybrid sketching idea brings these two extremes together and allows schemes that are a combination of the two extremes. 

In a distributed computing setting, the amount of data that can be fit into the memory of a worker node and the size of the largest problem that can be solved by that worker often do not match. The hybrid sketch idea is motivated by this mismatch and it is basically a sequentially concatenated sketching scheme where we first perform uniform sampling with dimension $m^\prime$ and then sketch the sampled data using another sketch preferably with better convergence properties (say, Gaussian) with dimension $m$. In other words, the worker reads $m^\prime$ rows of the data matrix $A$ and then applies sketching, reducing the number of rows from $m^\prime$ to $m$. 

Note that if $m^\prime = m$, hybrid sketch reduces to sampling; if $m^\prime = n$, then it reduces to Gaussian sketch; hence hybrid sketch can be thought of as middle ground between these. We note that this scheme does not take privacy into account as workers are assumed to have access to the original data.

We present experiment results in the numerical results section showing the practicality of the hybrid sketch idea. For the experiments involving very large-scale datasets, we have used Sparse Johnson-Lindenstrauss Transform (SJLT) \cite{nelson2013osnap} as the second sketching method in the hybrid sketch due to its low computational complexity.

% We now formalize the notion of hybrid sketching. Let $m^\prime$ denote the total number of rows read by the worker (sampled from the whole dataset) and $m$ denotes the final sketching size. This can be thought of as a concatenated (sequentially) sketching scheme where we first use uniform sampling with dimension $m^\prime$ and another sketching method (say, SJLT) with dimension $m$. Note that if $m^\prime = m$, this reduces to sampling; if $m^\prime = n$, then it reduces to SJLT. Any value between $m$ and $n$ gives a concatenated sketching scheme.

%We claim that it leads to better performance to read as much data as possible and then reduce the dimension to $n_{ls}$ via a sketching matrix that has better convergence than uniform sampling. 

%This idea can be thought of as offering trade-off between sketching speed and convergence rate (as a funct.

%Let us also assume that each worker is able to process $n_{p}$ rows of data where $n_{p} \geq n_{ls}$. In this case, one option is to read $n_{ls}$ rows of data into memory and solve the least squares problem, and this would be the fastest scenario. Another option is to read $n_{p}$ rows of data into memory and apply sketching to it to reduce the dimension to $n_{ls}$. 

% \todo[inline]{TODO: Theoretical result for hybrid sketch.}
\section{Distributed Sketching for Least-Norm Problems}
\label{sec:right_sketch}
Now we consider the high dimensional case where $n<d$ and applying the sketching matrix from the right (right-sketch), i.e., on the features. Let us define the minimum norm solution
\begin{align}
     x^* = &\arg\min_x \|x\|_2^2 \quad \mbox{s.t.} ~ Ax = b. 
 \end{align}
The solution to the above problem has a closed-form solution given by $x^* = A^T (AA^T)^{-1}b$ when the matrix $A$ is full row rank. We will assume that the full row rank condition holds in the sequel. Let us denote the optimal value of the minimum norm objective as $f(x^*)=\|x^*\|_2^2 = b^T (AA^T)^{-1} b$. Then we consider the approximate solution given by $\hat x_k = S_k^T \hat{z}_k$ and $\hat{z}_k$ is given by
\begin{align}
    \hat{z}_k = &\arg \min_z \|z\|_2^2 \quad \mbox{s.t.} ~ AS_k^T z = b,
\end{align}
where $S_k\in\real^{m\times d}$ and $z \in \real^m$. The averaged solution is computed as $\bar{x}=\frac{1}{q}\sum_{k=1}^q \hat{x}_k$.
%%% removed algorithm 2 to save space %%%
%The algorithm is listed in Alg. \ref{dist_rightsketch_alg}.
% \DontPrintSemicolon
% \begin{algorithm}
%  \KwIn{Data matrix $A \in \mathbb{R}^{n\times d}$, target vector $b \in \mathbb{R}^{n}$.}
%  \textbf{Workers $k=1,...,q$ in parallel: } \;
%  Sample $S_k \in \mathbb{R}^{m\times d}$. \;
%  Compute sketched data $AS_k^T$. \;
%  Solve $\hat{z}_k = \arg\min \|z\|_2^2$ subject to $ AS_k^Tz=b$. \;
%  Compute $\hat{x}_k = S_k^T \hat{z}_k$ and send to master. \;
%  \textbf{Master:} \;
%  return $\bar x \defn \frac{1}{q} \sum_{k=1}^q \hat x_k$. \;
%  \caption{Distributed right sketch for $n<d$ case.}
%  \label{dist_rightsketch_alg}
% \end{algorithm}
%\subsection{Gaussian Sketch}
Now we consider sketching matrices that are i.i.d. Gaussian. Lemma \ref{gaussian_one_sketch_rightsketch} establishes that the right-sketch estimator is an unbiased estimator and gives the exact expression for the expectation of the approximation error.
\blems \label{gaussian_one_sketch_rightsketch}
For the Gaussian sketch with sketch size satisfying $m>n+1$, the estimator $\hat{x}_k$ satisfies
\begin{align*}
    \Exs [\| \hat x_k - x^* \|_2^2] = \frac{d-n}{m-n-1} f(x^*).
\end{align*}
\elems
An exact formula for averaging multiple sketches that parallels Theorem \ref{Thm:AvgGaussian} can be obtained in a similar fashion. We defer the details to the supplement.

\section{Numerical Results}
We have implemented our distributed sketching methods for AWS Lambda in Python using Pywren \cite{jonas2017pywren}, which is a framework for serverless computing. The setting for the experiments is a centralized computing model where a single master node collects and averages the outputs of the $q$ worker nodes. Most of the figures in this section plot the approximation error which is computed by $(f(\bar{x})-f(x^*)) / f(x^*)$.

\subsection{Airline Dataset}
We have conducted experiments in the publicly available Airline dataset \cite{airline_dataset}. This dataset contains information on domestic USA flights between the years 1987-2008. We are interested in predicting whether there is going to be a departure delay or not, based on information about the flights. %The dataset contains information for around $120$ million flights with each flight having $29$ attributes.
More precisely, we are interested in predicting whether \texttt{DepDelay} $>15$ minutes using the attributes \texttt{Month}, \texttt{DayofMonth}, \texttt{DayofWeek}, \texttt{CRSDepTime}, \texttt{CRSElapsedTime}, \texttt{Dest}, \texttt{Origin}, and \texttt{Distance}. Most of these attributes are categorical and we have used dummy coding to convert these categorical attributes into binary representations. The size of the input matrix $A$, after converting categorical features into binary representations, becomes $(1.21 \times 10^8) \times 774$.%, and the output vector $b$ is a $(1.21 \times 10^8)$-dimensional vector.

%The aim of the experiments presented in this subsection is not to address the predictive power of linear least squares in predicting flight delays. Instead, 

%The goal of the experiments presented in this subsection is to illustrate the ability of distributed sketching to solve large-scale linear least squares problems. Even though solving a linear least squares problem by itself does not lead to high accuracy predictions, many other approaches to this problem requires solving least squares problems as a step. 
%Hence, it is of significance to have efficient tools for solving large-scale linear least squares problems with convergence guarantees.

\begin{figure}[htb]
%\vspace*{-1mm}
\begin{minipage}[b]{0.24\linewidth}
  \centering
  \centerline{\includegraphics[width=\columnwidth]{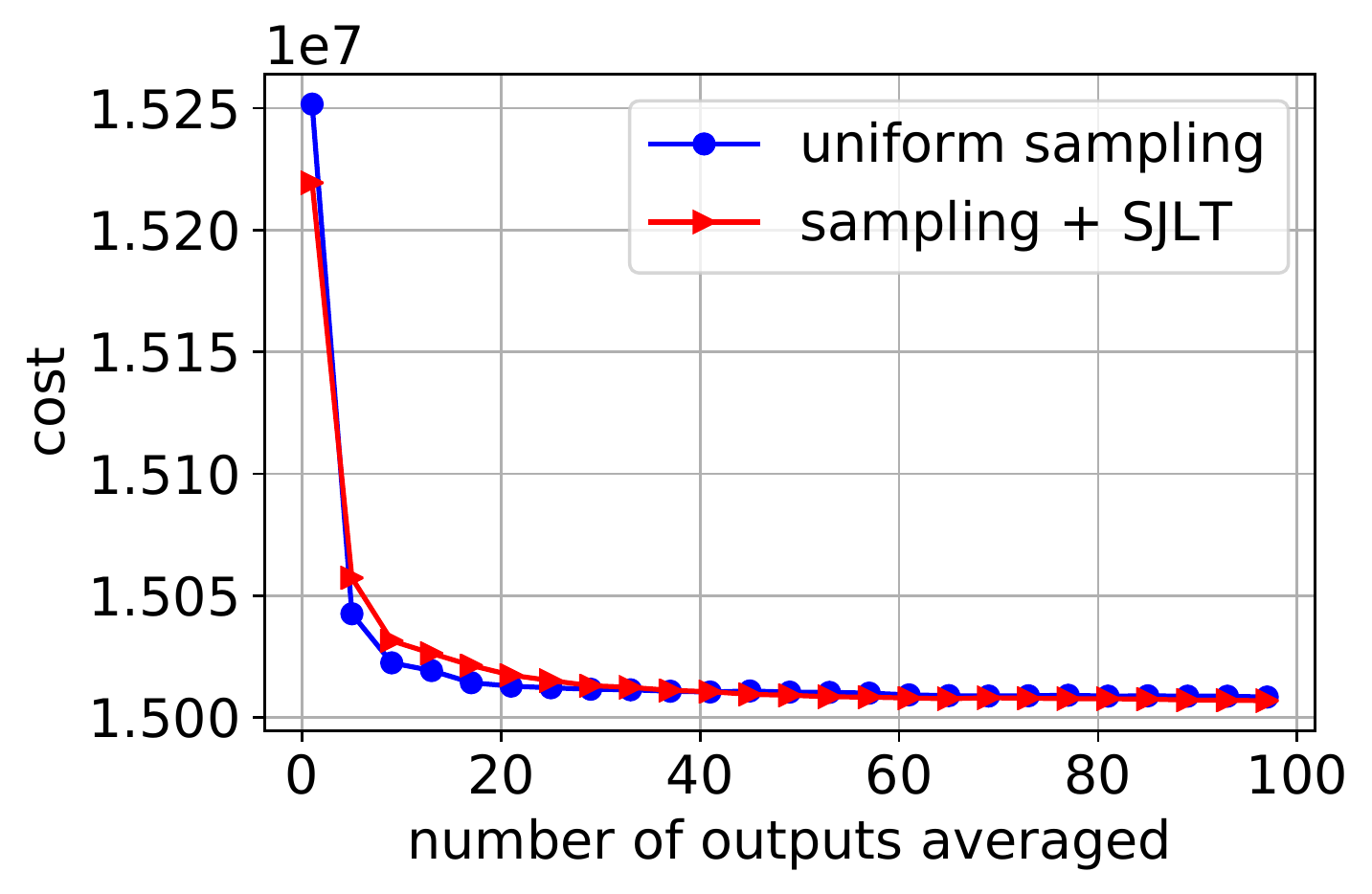}}
%  \vspace{2.0cm}
%\vspace*{-2mm}
  \centerline{a) $m^\prime = 5m=5\times 10^5$}\medskip
\end{minipage}
\hfill
\begin{minipage}[b]{0.24\linewidth}
  \centering
  \centerline{\includegraphics[width=\columnwidth]{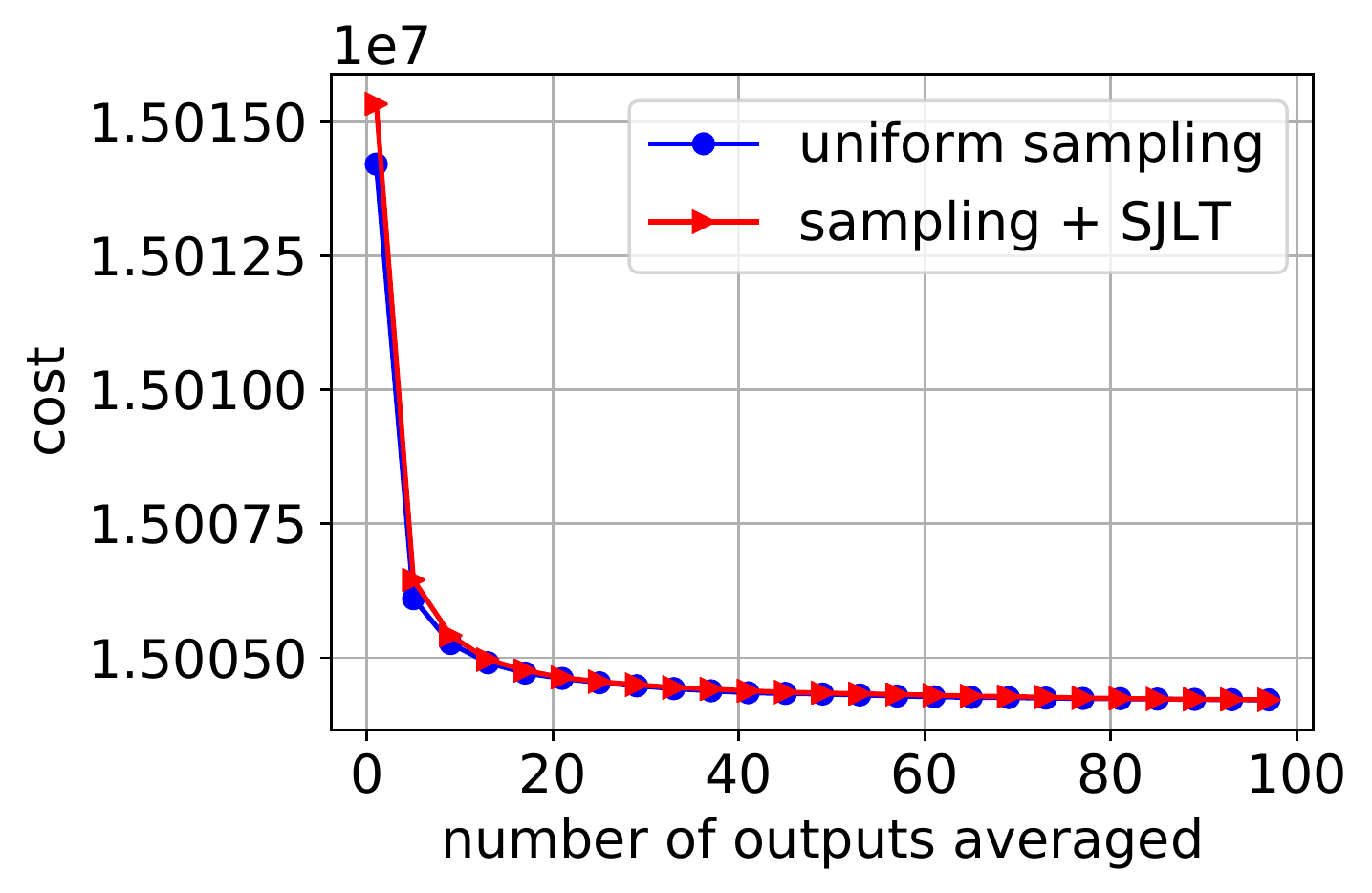}}
%  \vspace{2.0cm}
%\vspace*{-2mm}
  \centerline{b) $m^\prime=2m=2\times 10^6$}\medskip
\end{minipage}
\hfill
\begin{minipage}[b]{0.24\linewidth}
  \centering
  \centerline{\includegraphics[width=\columnwidth]{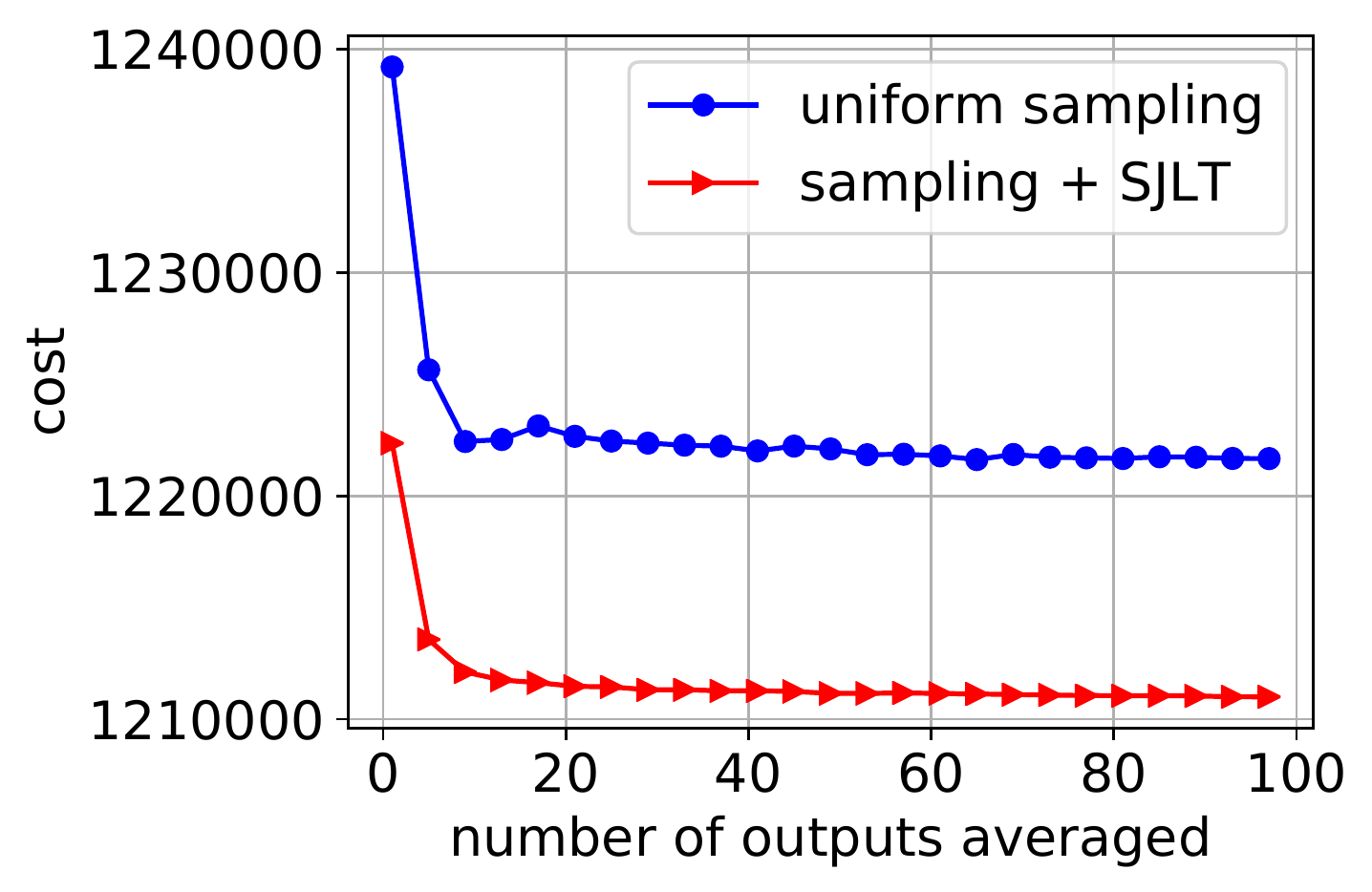}}
%  \vspace{1.5cm}
%\vspace*{-2mm}
  \centerline{c) $m^\prime = 5m=5\times 10^5$}\medskip
\end{minipage}
\hfill
\begin{minipage}[b]{0.24\linewidth}
  \centering
  \centerline{\includegraphics[width=\columnwidth]{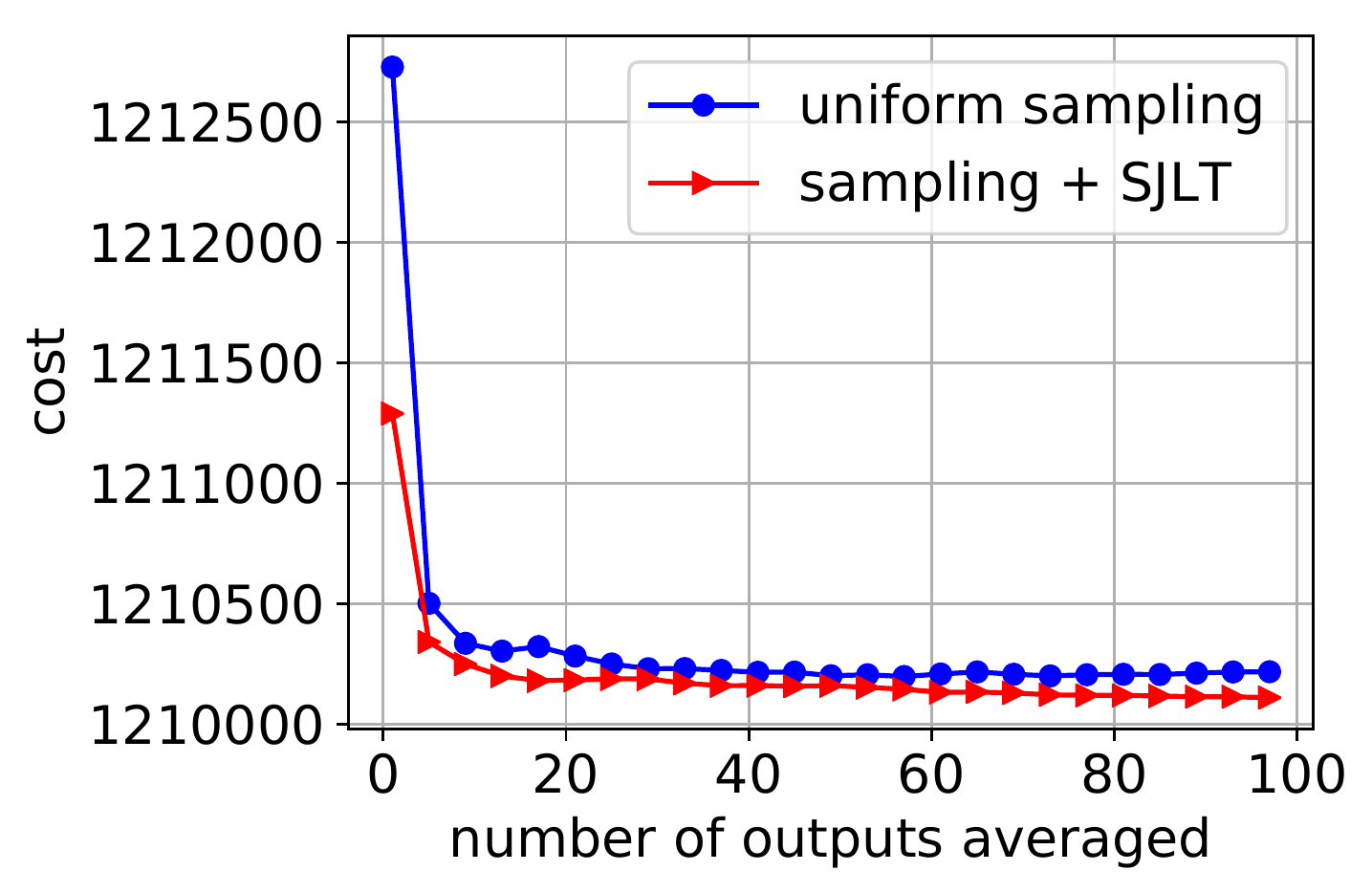}}
%  \vspace{2.0cm}
%\vspace*{-2mm}
  \centerline{d) $m^\prime = 2m=2\times 10^6$}\medskip
\end{minipage}
\vspace*{-2mm}
\caption{AWS Lambda experiments on the entire airline dataset ($n=1.21\times 10^8$) with $q=100$ workers. %The vertical axis represents the approximation error, and the horizontal axis shows the number of outputs used in obtaining the averaged estimate. 
The run times for each plot are as follows (given in this order: sampling, sampling followed by SJLT): a: 37.5, 43.9 seconds, b: 48.3, 60.1 seconds, c: 39.8, 52.9 seconds, d: 78.8, 107.6 seconds.}
\label{airline_cloud}
%\vspace*{-2mm}
\end{figure}

We have solved the linear least squares problem on the entire airline dataset: $\text{minimize}_x \|Ax-b\|_2^2$ using $q$ workers on AWS Lambda. The output $b$ for the plots a and b in Fig. \ref{airline_cloud} is a vector of binary variables indicating delay. The output $b$ for the plots c and d in Fig. \ref{airline_cloud} is artificially generated via $b=Ax_{truth}+ \epsilon$ where $x_{truth}$ is the underlying solution and $\epsilon$ is random Gaussian noise distributed as $\mathcal{N}(0, 0.01I)$. Fig. \ref{airline_cloud} shows that sampling followed by SJLT leads to a lower error.

%Fig. \ref{airline_cloud} shows the results for the entire dataset where we have used AWS Lambda functions in parallel as the computing platform. 
Note that it makes sense to choose $m$ and $m^\prime$ as large as the available resources allow because for larger $m$ and $m^\prime$, the convergence is faster. Based on the run times given in the caption of Fig. \ref{airline_cloud} we see that workers take longer to finish their tasks if SJLT is involved. Decreasing $m^\prime$ will help reduce this processing time at the expense of error performance.

% \subsubsection{Nonlinear Least Squares using Gauss-Newton Method}
% We now pose a different problem that we then solve using the proposed mechanism. We are now interested in predicting whether there is going to be a delay or not. So, we solve the following classification problem:
% \begin{align}
%     \text{minimize}_{x} \|g(Ax)-y\|_2^2
% \end{align}
% where $g(.)$ is the sigmoid function, applied element-wise if its argument is a vector. $y_i=0$ if there is no delay and $y_i=1$ if there is delay in the $i$'th flight. A flight is considered delayed if it arrives at least 15 minutes late.

% To solve this problem, we have used the Gauss-Newton method where for each update, we use the averaged sketching method on serverless AWS Lambda functions to solve the resulting large-scale least squares problem 
% \begin{align}
%     x^{k+1} = \arg \min \|D^kx-b^k\|_2^2
% \end{align}
% where $D^k$ denotes the Jacobian matrix at the $k$'th iteration, that is, $D^k = diag(\frac{\exp(-\tilde{a}_i^T x^k)}{(1+\exp(-\tilde{a}_i^T x^k))^2}) A$ and $b^k=D^kx^k-g(Ax^k)+y$. Note that $\tilde{a}_i^T$ denotes the $i$'th row of $A$.

% Fig. ... illustrates the ...

\subsection{Image Dataset: Extended MNIST}
The experiments of this subsection are performed on the image dataset EMNIST (extended MNIST) \cite{emnist_dataset}. We have used the "bymerge" split of EMNIST, which has 700K training and 115K test images. The dimensions of the images are $28\times 28$ and there are 47 classes in total (letters and digits). %Some capital letter classes have been merged with small letter classes (like C and c), that is why we have 47 classes and not 62.

\begin{figure}[htb]
%\vspace*{-1mm}
\begin{minipage}[b]{0.33\linewidth}
  \centering
  \centerline{\includegraphics[width=\columnwidth]{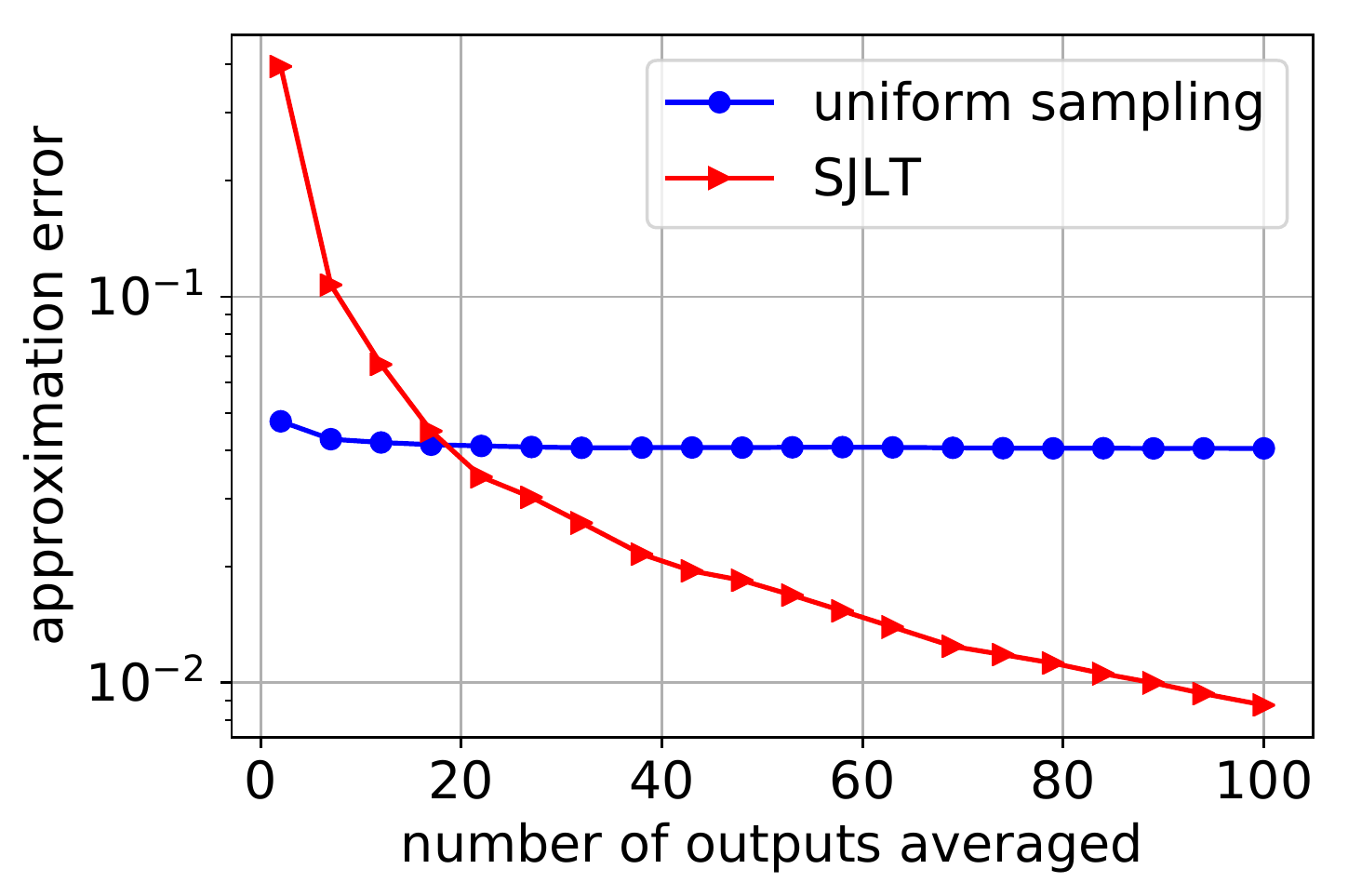}}
%  \vspace{2.0cm}
%\vspace*{-2mm}
  \centerline{a) Approximation error}\medskip
\end{minipage}
\hfill
\begin{minipage}[b]{0.33\linewidth}
  \centering
  \centerline{\includegraphics[width=\columnwidth]{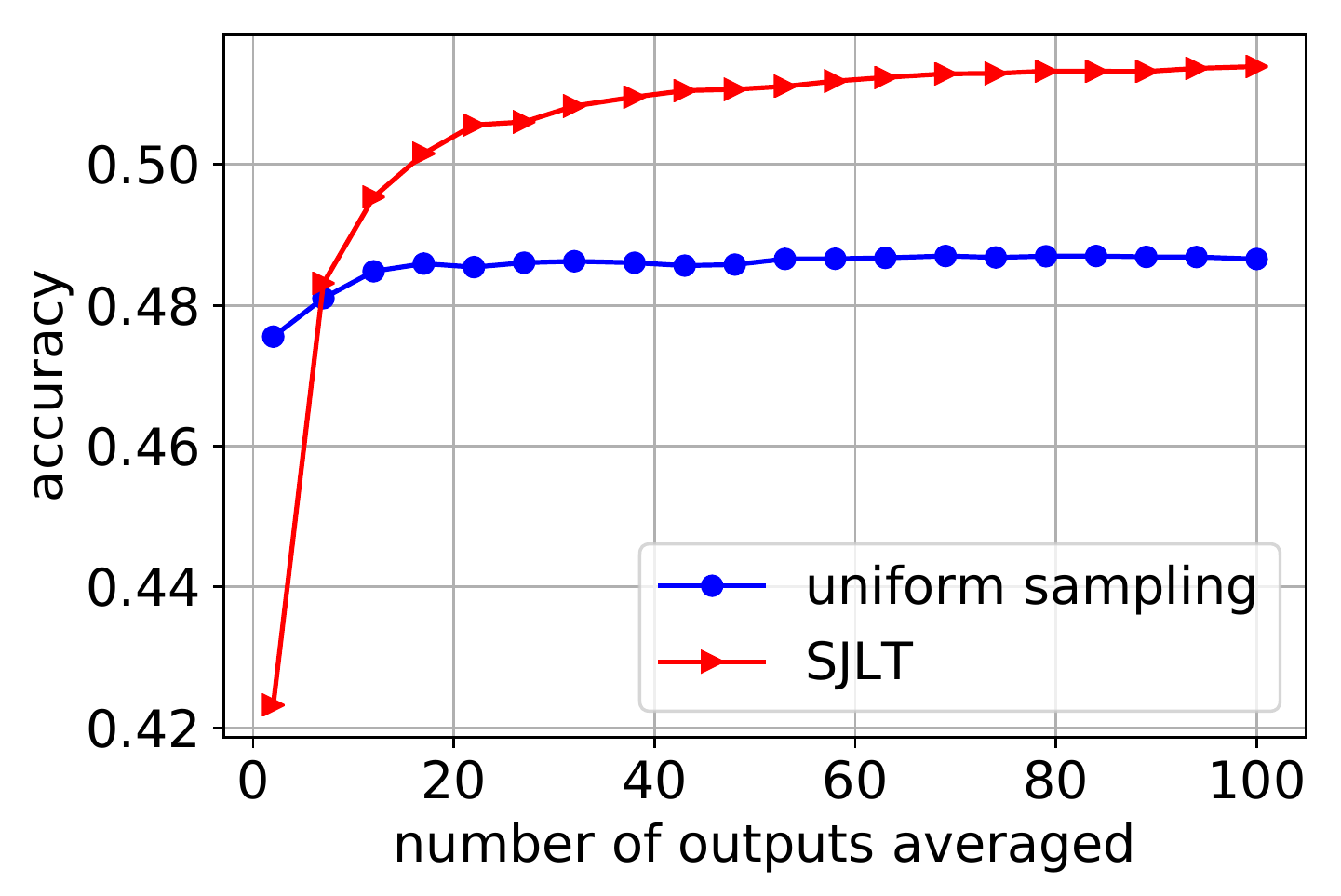}}
%  \vspace{2.0cm}
%\vspace*{-2mm}
  \centerline{b) Test accuracy}\medskip
\end{minipage}
\vspace*{-2mm}
\caption{Approximation error and test accuracy plots for the EMNIST-bymerge dataset where $q=100$, $m=2000$, $s=20$. The run times on AWS Lambda for uniform sampling and SJLT are $41.5$ and $66.9$ seconds, respectively.}
\label{emnist_experiments}
%\vspace*{-2mm}
\end{figure}

% \begin{figure}[htb]
% \begin{minipage}[b]{0.48\linewidth}
%   \centering
%   \centerline{\includegraphics[width=4.5cm]{src/emnist_cost_sampling.pdf}}
% %  \vspace{2.0cm}
%   \centerline{(a) Cost, sampling}\medskip
% \end{minipage}
% \hfill
% \begin{minipage}[b]{0.48\linewidth}
%   \centering
%   \centerline{\includegraphics[width=4.5cm]{src/emnist_cost_sjlt.pdf}}
% %  \vspace{2.0cm}
%   \centerline{(b) Cost, SJLT}\medskip
% \end{minipage}
% \hfill
% \begin{minipage}[b]{.48\linewidth}
%   \centering
%   \centerline{\includegraphics[width=4.5cm]{src/emnist_test_accuracy_sampling.pdf}}
% %  \vspace{1.5cm}
%   \centerline{(c) Test accuracy, sampling}\medskip
% \end{minipage}
% \hfill
% \begin{minipage}[b]{0.48\linewidth}
%   \centering
%   \centerline{\includegraphics[width=4.5cm]{src/emnist_test_accuracy_sjlt.pdf}}
% %  \vspace{2.0cm}
%   \centerline{(d) Test accuracy, SJLT}\medskip
% \end{minipage}
% %\vspace{-0.5cm}
% \caption{Cost and test accuracy plots for EMNIST-bymerge dataset.}
% \label{emnist_experiments}
% \end{figure}

Fig. \ref{emnist_experiments} shows the approximation error and test accuracy plots when we solve the least squares problem on the EMNIST-bymerge dataset using the model averaging method. Because this is a multi-class classification problem, we have one-hot encoded the labels. %For these experiments, we have used $q=100$ workers in AWS Lambda and the sketch dimension is $m=2000$. 
Fig. \ref{emnist_experiments} demonstrates that SJLT is able to drive the cost down more and increase the accuracy more than uniform sampling.

\subsection{Performance on Large-Scale Synthetic Datasets}
\begin{figure}[htb]
%\vspace*{-1mm}
\begin{minipage}[b]{0.33\linewidth}
  \centering
  \centerline{\includegraphics[width=\columnwidth]{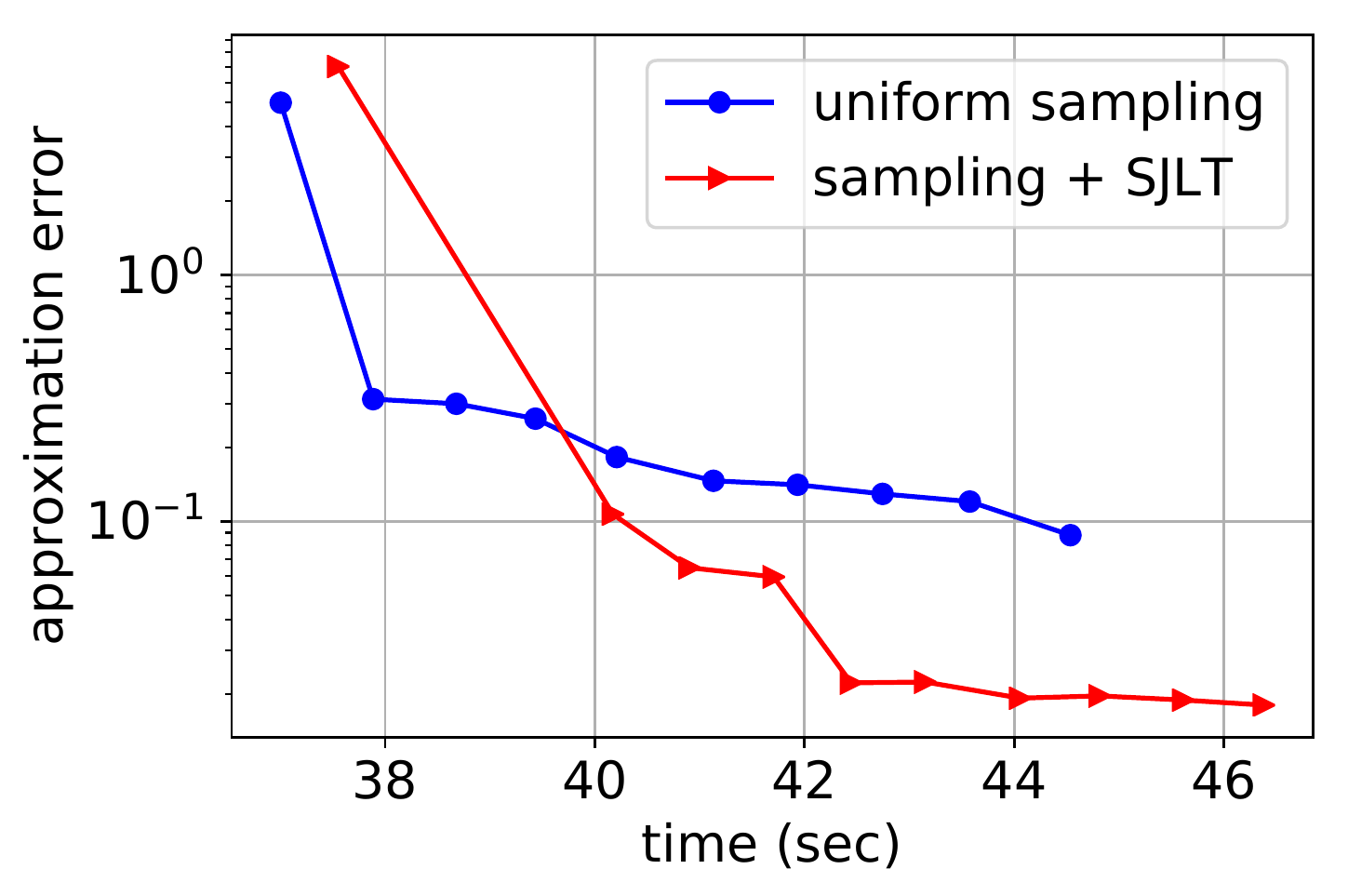}}
%  \vspace{2.0cm}
%\vspace*{-2mm}
  \centerline{a) $A\in \mathbb{R}^{10^7 \times 10^3}$}\medskip
\end{minipage}
\hfill
\begin{minipage}[b]{0.33\linewidth}
  \centering
  \centerline{\includegraphics[width=\columnwidth]{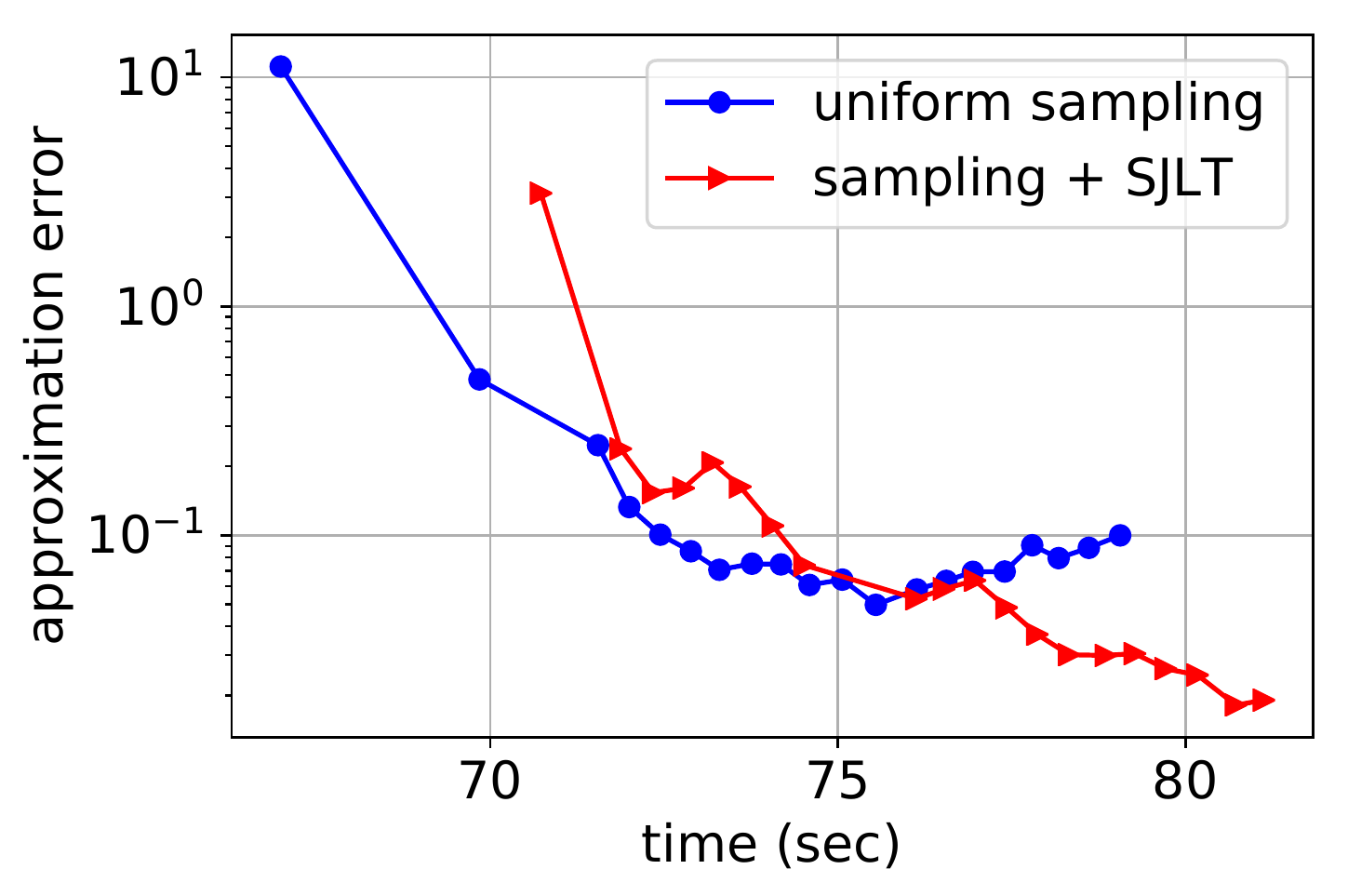}}
%  \vspace{2.0cm}
  \centerline{b) $A\in \mathbb{R}^{(2\times 10^7) \times (2\times 10^3)}$}\medskip
\end{minipage}
% \hfill
% \begin{minipage}[b]{0.48\linewidth}
%   \centering
%   \centerline{\includegraphics[width=4.5cm]{src/test_loss_1.pdf}}
% %  \vspace{2.0cm}
%   \centerline{(c) Test set 1}\medskip
% \end{minipage}
% \hfill
% \begin{minipage}[b]{0.48\linewidth}
%   \centering
%   \centerline{\includegraphics[width=4.5cm]{src/test_loss_2.pdf}}
% %  \vspace{2.0cm}
%   \centerline{(d) Test set 2}\medskip
% \end{minipage}
\vspace*{-2mm}
\caption{Approximation error vs time for AWS Lambda experiments on randomly generated large-scale datasets ($q=200$ AWS Lambda functions have been used). %Plots a, b show the approximation errors on the training sets and plots c, d show the test set errors.
}
\label{random_data_experiments}
%\vspace*{-2mm}
\end{figure}

This subsection contains the experiments performed on randomly generated large-scale data to illustrate scalability of the methods. Plots in Fig. \ref{random_data_experiments} show the approximation error as a function of time, where the problem dimensions are as follows: $A\in \mathbb{R}^{10^7 \times 10^3}$ for plot a and $A\in \mathbb{R}^{(2\times 10^7) \times (2\times 10^3)}$ for plot b. These data matrices take up $75$ GB and $298$ GB, respectively. The data used in these experiments were randomly generated from the student's t-distribution with degrees of freedom of $1.5$ for plot a and $1.7$ for plot b. The output vector $b$ was computed according to $b = A x_{truth} + \epsilon$ where $\epsilon \in \mathbb{R}^n$ is i.i.d. noise distributed as $\mathcal{N}(0, 0.1 I_n)$. Other parameters used in the experiments are $m=10^4, m^\prime = 10^5$ for plot a, and $m=8\times 10^3, m^\prime = 8\times 10^4$ for plot b. %Furthermore, plots c and d in Fig. \ref{random_data_experiments} show the approximation error on the test set. The test sets have also been randomly sampled from the same distribution as the training set and we have generated $10^6$ samples in both cases.
We observe that both plots in Fig. \ref{random_data_experiments} reveal similar trends where the hybrid approach leads to a lower approximation error but takes longer due to the additional processing required for SJLT. %We also observe that the curve for the hybrid approach is below the curve for sampling, that is, the average due to the hybrid approach is a better approximate than the one due to sampling most of the time. 

\subsection{Numerical Results for Right Sketch: $n<d$ Case}
\begin{figure}[htb]
%\vspace*{-1mm}
\begin{minipage}[b]{0.33\linewidth}
  \centering
  \centerline{\includegraphics[width=\columnwidth]{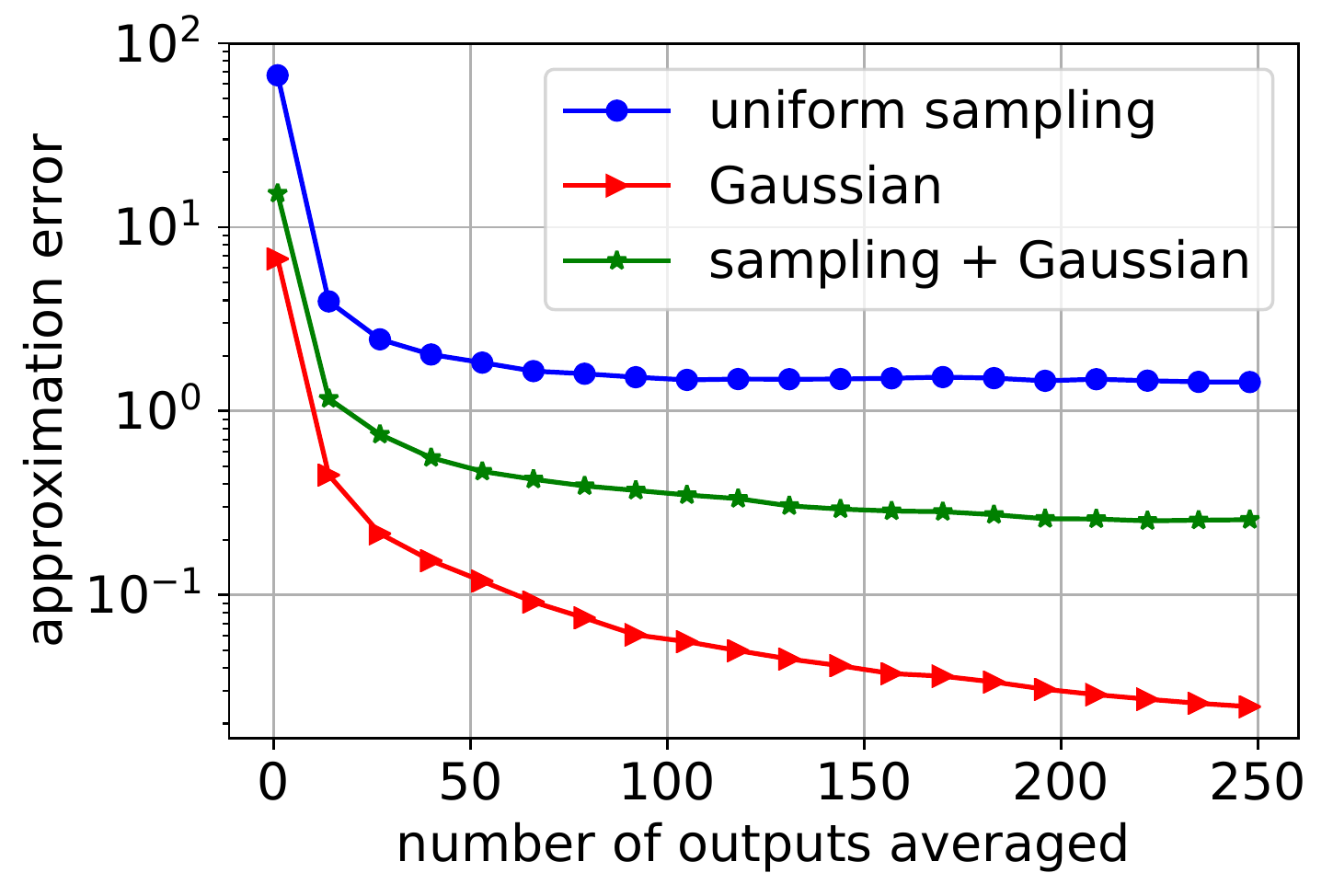}}
  %\vspace*{-2mm}
  \centerline{a) Random data}\medskip
\end{minipage}
\hfill
\begin{minipage}[b]{0.33\linewidth}
  \centering
  \centerline{\includegraphics[width=\columnwidth]{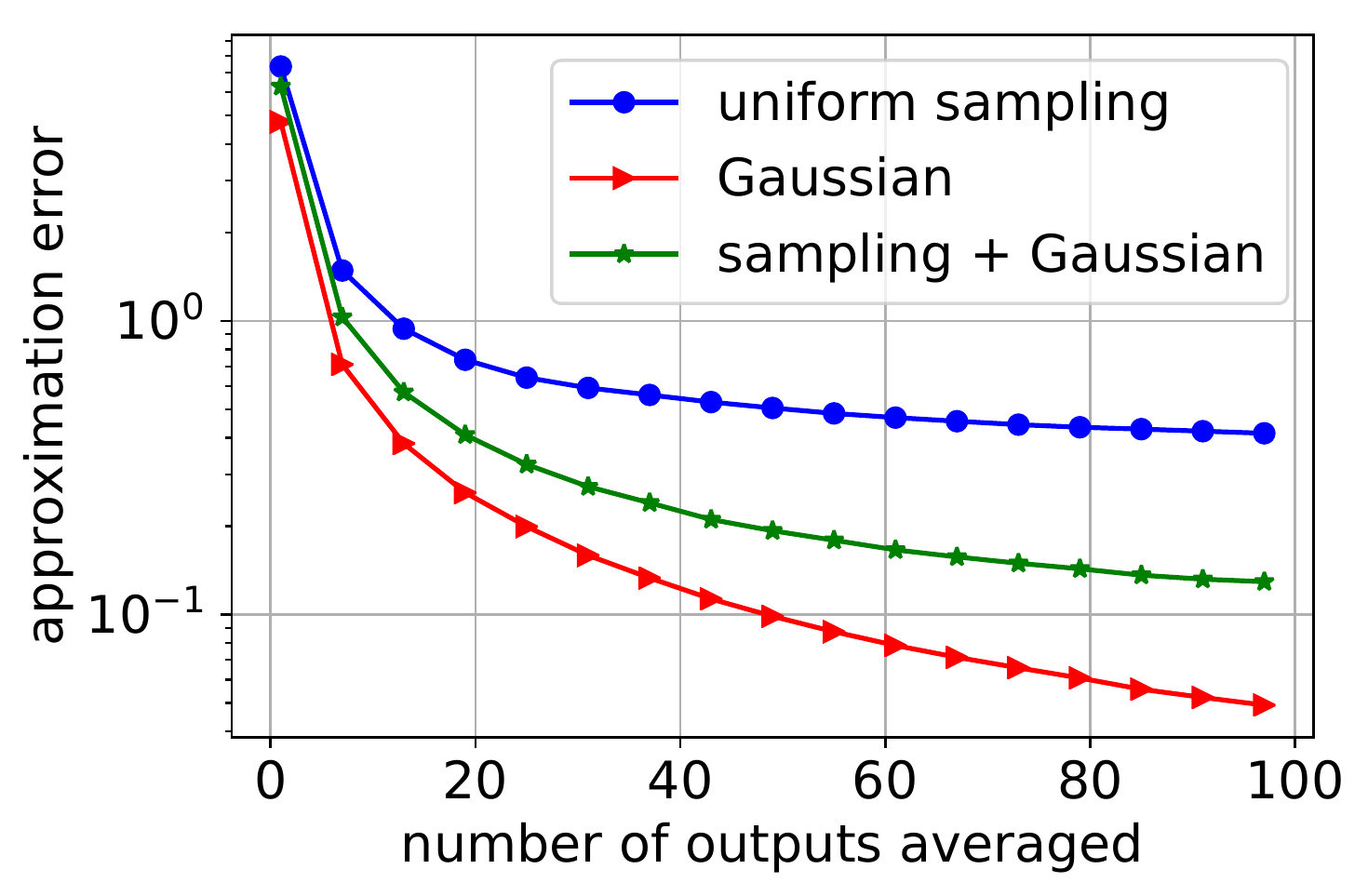}}
  %\vspace*{-2mm}
  \centerline{b) Airline data}\medskip
\end{minipage}
\vspace*{-2mm}
\caption{Averaging for least norm problems. \textit{Plot (a):} The parameters are $n=50$, $d=1000$, $m=200$, $m^\prime = 500$. \textit{Plot (b):} Least norm averaging applied to a subset of the airline dataset. The parameters are $n=2000$, $d=11588$, $m=4000$, $m^\prime =8000$. For this plot, the features include the pairwise interactions in addition to the original features.}
\label{right_sketch_plots}
%\vspace*{-2mm}
\end{figure}

Fig. \ref{right_sketch_plots} shows the approximation error as a function of the number of averaged outputs in solving the least norm problem for two different datasets. %The approximation error is the ratio of the $\ell_2$-norms of the averaged solution $\bar{x}$ and the optimal solution $x^*$, that is $\|\bar{x}\|_2^2 / \|x^*\|_2^2 - 1$.
The dataset for Fig. \ref{right_sketch_plots}(a) is randomly generated with dimensions $A \in \mathbb{R}^{50\times 1000}$. We observe that Gaussian sketch outperforms uniform sampling in terms of the approximation error. Furthermore, Fig. \ref{right_sketch_plots}(a) verifies that if we apply the hybrid approach of first sampling and then using Gaussian sketch, then its performance falls between the extreme ends of only sampling and only using Gaussian sketch. Moreover, Fig. \ref{right_sketch_plots}(b) shows the results for the same experiment on a subset of the airline dataset where we have included the pairwise interactions as features which makes this an underdetermined linear system. Originally, we had $774$ features for this dataset, if we include all $x_ix_j$ terms as features, we would have a total of $299925$ features, most of which are zero for all samples. We have excluded the all-zero columns from this extended matrix to obtain the final dimensions $2000\times 11588$. %This is to make sure that the sketched sub-problems are underdetermined as well. 

\section{Conclusion}
In this work, we have studied averaging sketched solutions for linear least squares problems for both $n>d$ and $n<d$. We have discussed distributed sketching methods from the perspectives of convergence, bias, privacy, and (serverless) computing framework. Our results and numerical experiments suggest that distributed sketching methods offer a competitive straggler-resilient solution for solving large scale linear least squares problems for distributed systems. 

\section*{Broader Impact}
This work does not present any foreseeable negative societal consequence. The methods presented are applicable to any type of data and could be useful for both practitioners and researchers whose work involve large scale scientific computing. The main advantages of the methods are privacy preservation and faster computing times which could be beneficial for many engineering and science applications.

% Authors are required to include a statement of the broader impact of their work, including its ethical aspects and future societal consequences. 
% Authors should discuss both positive and negative outcomes, if any. For instance, authors should discuss a) 
% who may benefit from this research, b) who may be put at disadvantage from this research, c) what are the consequences of failure of the system, and d) whether the task/method leverages
% biases in the data. If authors believe this is not applicable to them, authors can simply state this.

% Use unnumbered first level headings for this section, which should go at the end of the paper. {\bf Note that this section does not count towards the eight pages of content that are allowed.}
% \input{acknowledgements.tex}

\bibliographystyle{plain}
\bibliography{mert}

\newpage
\appendix
\section{Additional Numerical Results}
In the numerical results section of the main paper, we have presented large scale experiments that have been run on the AWS Lambda platform. In the case of large datasets and limited computing resources such as memory and lifetime, most of the standard sketches are computationally too expensive as discussed in the main body of the paper. This is the reason why we limited the scope of the large scale experiments to uniform sampling, SJLT, and hybrid sketch. In this section we present some additional experimental results on smaller datasets to empirically verify the theoretical results of the paper.

\begin{figure}[ht]
\begin{minipage}[b]{0.48\linewidth}
  \centering
  \centerline{\includegraphics[width=\columnwidth]{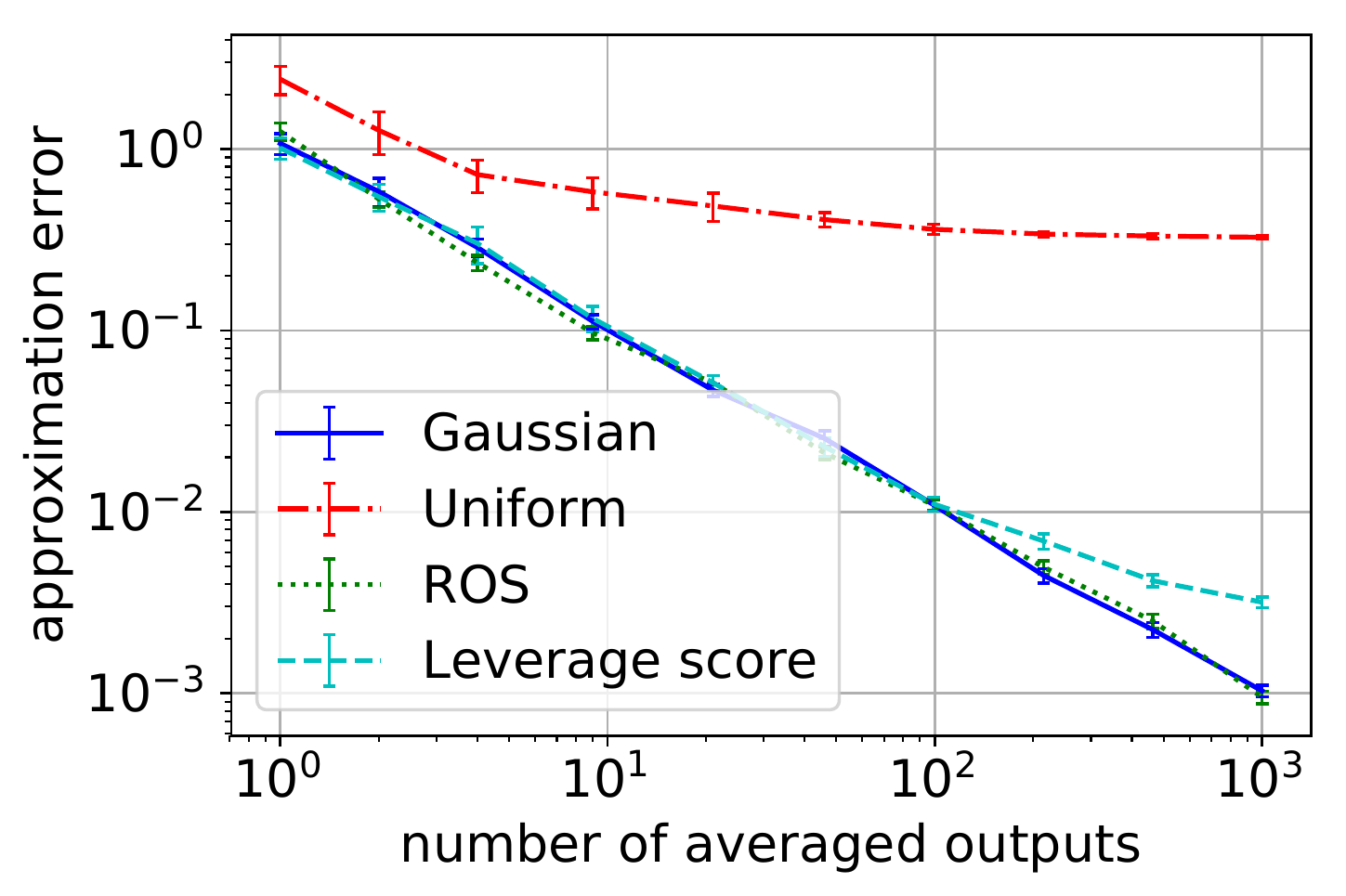}}
  \centerline{a) echocardiogram}\medskip
\end{minipage}
\hfill
\begin{minipage}[b]{0.48\linewidth}
  \centering
  \centerline{\includegraphics[width=\columnwidth]{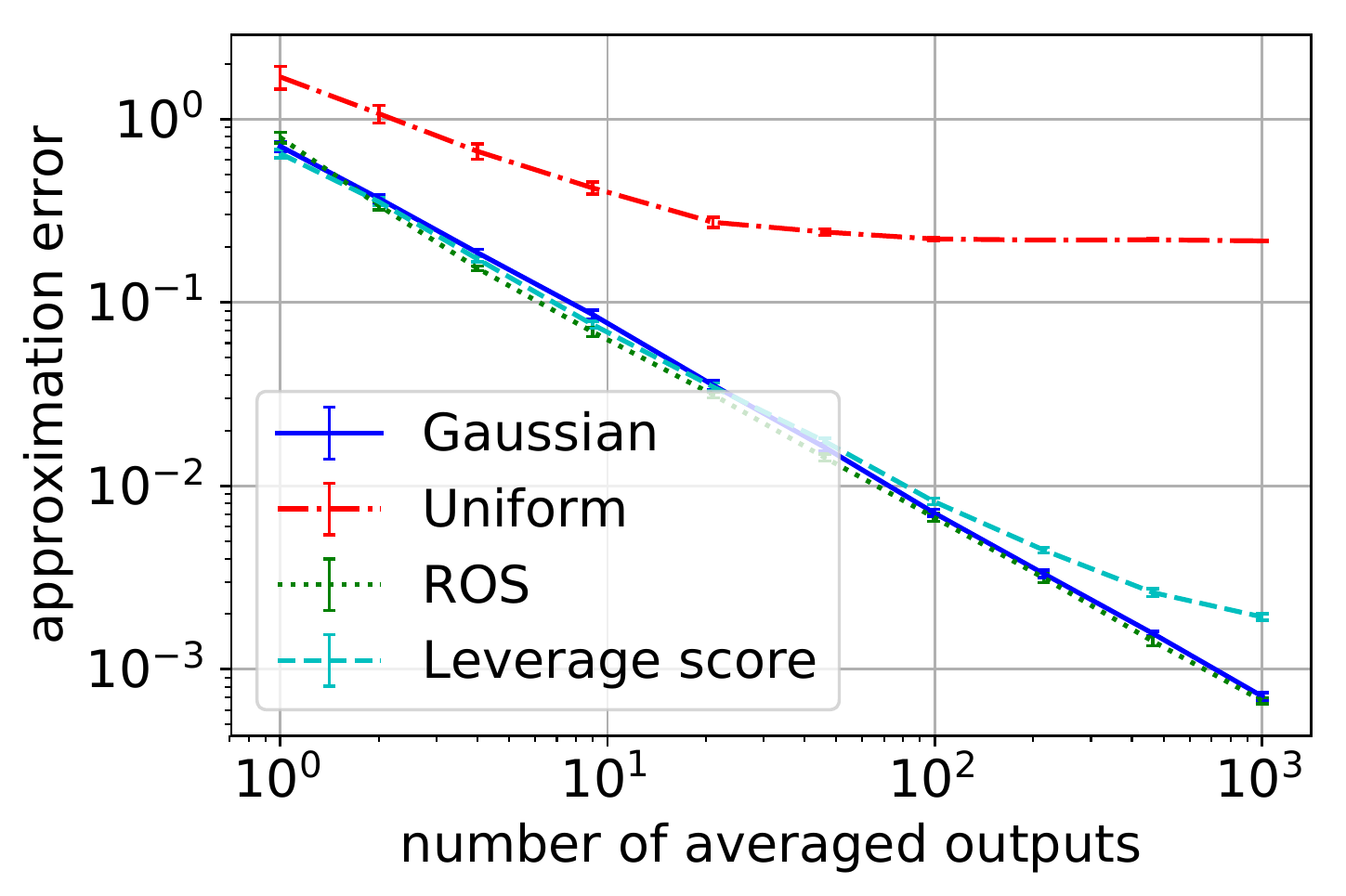}}
  \centerline{b) oocytes merluccius nucleus 4d}\medskip
\end{minipage}
\caption{Approximation error against the number of averaged outputs in log-log scale for various sketching methods on UCI datasets. All of the curves have been averaged over $25$ independent trials and the vertical error bars show the standard error. The parameters used are as follows. Plot a: $n=131$, $d=10$, $m=20$. Plot b: $n=1022$, $d=41$, $m=100$.}
\label{supplement_uci_datasets}
\end{figure}

We present results on two UCI datasets in Figure \ref{supplement_uci_datasets} comparing the performances of the sketches we discussed in the paper.

Figure \ref{supplement_uci_datasets} shows that Gaussian and ROS sketches lead to unbiased estimators in the experiments because the corresponding curves appear linear in the log-log scale plots. The experimental results suggest that the upper bound that we have found for the bias of the ROS sketch may not be tight. We see that the estimates for uniform sampling and leverage score sampling are biased. The observation that the Gaussian sketch estimator is unbiased in the experiments is perfectly consistent with our theoretical findings. Furthermore, we observe that the approximation error is the highest in uniform sampling, which is also in agreement with the theoretical upper bounds that we have presented.
\section{Proofs}

\subsection{Gaussian Sketch and Other Sketches}
We give below the proofs for the rest of the lemmas and theorems stated in the paper.

\proof[\textbf{Proof of Lemma \ref{gaussian_one_sketch}}]
Suppose that the matrix $A$ is full column rank. Then, for $m\ge d$, the matrix $A^T S_k^T S_k A$ follows a Wishart distribution, and is invertible with probability one. Conditioned on the invertibility of $A^T S_k^T S_k A$, we have 
\begin{align*}
    \hat x_k = (A^T S_k^T S_k A)^{-1} A^T S_k^T S_k b &= (A^T S_k^T S_k A)^{-1} A^T S_k^T S_k(Ax^* + b^\perp)\\
             &= x^* + (A^T S_k^T S_k A)^{-1} A^T S_k^T S_k b^\perp\,,
\end{align*}
where we have defined $b^\perp = b-Ax^*$\,.
Note that $S_kA$ and $S_k b^\perp$ are independent random matrices since they are Gaussian and uncorrelated as a result of the normal equations $A^T b^\perp = 0$.
Conditioned on the realization of the matrix $S_kA$ and the event $A^T S^T_kS_k A \succ 0$, a simple covariance calculation shows that
\begin{align}
    \hat x_k \sim \mathcal{N}\big( x^*, \frac{1}{m} f(x^*) (A^T S_k^T S_k A)^{-1}\big)\,.\nonumber
\end{align}
Multiplying with the data matrix $A$ on the left yields the distribution of the prediction error, conditioned on $S_kA$, as
\begin{align}
    A(\hat x_k-x^*) \sim \mathcal{N}\big( 0,  \frac{1}{m} f(x^*) A(A^T S_k^T S_k A)^{-1}A^T\big)\,.\nonumber
\end{align}
Then we can compute the conditional expectation of the squared norm of the error
\begin{align}
    \Exs [ \| A(\hat x_k - x^*) \|_2^2 ~\, \big \vert  S_kA ] = \frac{f(x^*)}{m} \Exs [\mathrm{tr}  A(A^T S_k^T S_k A)^{-1}A^T] \,.\nonumber
\end{align}
Next we recall that the expected inverse of the Wishart matrix $A^T S_k^T S_k A$ satisfies (see, e.g.,\cite{letac2004all})
\begin{align}
    \Exs [(A^T S_k^T S_k A)^{-1}] = (A^T A)^{-1} \frac{m}{m-d-1}\,.\nonumber
\end{align}
Plugging in the previous result, using the tower property of expectations and noting that $\mbox{tr} A (A^TA)^{-1} A = \mbox{rank} A = d$, we obtain the claimed result.
\qed

\proof[\textbf{Proof of Theorem \ref{Thm:AvgGaussian}}]
% We first derive a result on the expected error of the averaged solution that relates the error of a single estimator $\hat{x}_k$ to that of the averaged solution $\bar{x}$.
% \begin{align*}
% \Exs &[\|A\bar x - Ax^*\|_2^2] = \Exs \left[ \left \|\frac{1}{q} \sum_{k=1}^q (A \hat x_k - \Exs[A \hat x_k]) \right\|_2^2 \right] \\
% & = \frac{1}{q^2}  \Exs \sum_{k=1}^q \sum_{l=1}^q \inprod{A \hat x_k - \Exs[A \hat x_k]}{A \hat x_l - \Exs[A  \hat x_l]}\\
% & = \frac{1}{q^2} \sum_{k=1}^q \Exs[\|A \hat x_k - \Exs [ A \hat x_k]\|^2]\\
% &= \frac{1}{q} \Exs[\|A \hat x_1 -  A \xstar]\|_2^2,
% \end{align*}
Because the Gaussian sketch estimator is unbiased (i.e., $\Exs[\hat{x}_k]=x^*$), Lemma \ref{expected_obj_val_diff} reduces to $
    \Exs[f(\bar{x})] - f(x^*) = \frac{1}{q} \Exs[\|A \hat x_1 -  A \xstar]\|_2^2.
$
By Lemma \ref{gaussian_one_sketch}, the error of the averaged solution conditioned on the events that $E_k = A^TS_k^TS_kA \succ 0$, $\forall i=1,...,q$ can exactly be written as
\begin{align*}
    \Exs[ \| A(\bar{x} - x^*) \|_2^2 | E_1 \cap ...\cap E_q ] = \frac{1}{q} \frac{d}{m-d-1} f(x^*).
\end{align*}
Using Markov's inequality, it follows that
\begin{align*}
    P(\| A(\bar{x} - x^*) \|_2^2 \geq a |  E_1 \cap ...\cap E_q) \leq \frac{1}{qa} \frac{d}{m-d-1} f(x^*).
\end{align*}
The LHS can be rewritten as 
\begin{align}
    \frac{P(\| A(\bar{x} - x^*) \|_2^2 \geq a \cap (\bigcap_{k=1}^q E_k))}{P(\bigcap_{k=1}^q E_k)} &\geq \frac{P(\| A(\bar{x} - x^*) \|_2^2 \geq a) + P(\bigcap_{k=1}^q E_k) - 1}{P(\bigcap_{k=1}^q E_k)} \label{temp_1}\\
    &= \frac{P(\| A(\bar{x} - x^*) \|_2^2 \geq a) + P(E_1)^q - 1}{P(E_1)^q}, \label{temp_2}
\end{align}
where we have used the identity $P(A\cap B)\geq P(A)+P(B)-1$ in \eqref{temp_1} and the independence of the events $E_k$ in \eqref{temp_2}. It follows
\begin{align*}
    P(\| A(\bar{x} - x^*) \|_2^2 \leq a) \geq P(E_1)^q \left(1 - \frac{1}{qa} \frac{d}{m-d-1} f(x^*) \right).
\end{align*}
Setting $a=f(x^*)\frac{\epsilon}{q}$ and plugging in $P(E_1)\geq 1-e^{-c_1m}$ where $c_1$ is a constant \cite{PilWai14b}, we obtain
\begin{align*}
    P\left(\frac{\| A(\bar{x} - x^*) \|_2^2}{f(x^*)} \leq \frac{\epsilon}{q}\right)
    \geq (1-e^{-c_1m})^q \left(1 - \frac{1}{\epsilon} \frac{d}{m-d-1} \right).
\end{align*}
\qed

\proof[\textbf{Proof of Lemma \ref{expected_obj_val_diff}}]
The expectation of the difference between $f(\bar{x})$ and $f(x^*)$ is given by:
\begin{align*}
    \Exs[f(\bar{x})]-f(x^*) &= \Exs [\|A(\bar{x}-x^*)\|_2^2] \\
    &= \Exs\left[ \left \|\frac{1}{q}\sum_{k=1}^q (A\hat{x}_k - Ax^*) \right \|_2^2 \right] \\
    &= \frac{1}{q^2} \Exs \left[ \sum_{k=1}^q\sum_{l=1}^q \langle A\hat{x}_k - Ax^*, A\hat{x}_l-Ax^* \rangle \right] \\
    &= \frac{1}{q^2} \sum_{k=1}^q \Exs \left[ \|A\hat{x}_k-Ax^* \|_2^2 \right] + \frac{1}{q^2} \sum_{k\neq l\,, 1 \leq k,l \leq q} \Exs \left[ \langle A\hat{x}_k - Ax^*, A\hat{x}_l-Ax^* \rangle \right] \\
    &= \frac{1}{q} \Exs \left[ \|A\hat{x}_1-Ax^* \|_2^2 \right] + \frac{q^2-q}{q^2} \Exs \left[ \langle A\hat{x}_1 - Ax^*, A\hat{x}_2-Ax^* \rangle \right] \\
    &= \frac{1}{q} \Exs \left[ \|A\hat{x}_1-Ax^* \|_2^2 \right] + \frac{q-1}{q} \Exs[A\hat{x}_1 - Ax^*]^T \Exs[A\hat{x}_2 - Ax^*] \\
    &= \frac{1}{q} \Exs \left[ \|A\hat{x}_1-Ax^* \|_2^2 \right] + \frac{q-1}{q} \| \Exs[A\hat{x}_1] - Ax^* \|_2^2,
\end{align*}
where the first line follows as in \eqref{eq:expected_diff}.
\qed

\begin{proof}[\textbf{Proof of Lemma \ref{norm_of_bias}}]
Assuming $A^TS_k^TS_kA$ is invertible (we will take the probability of this event into account later), the norm of the bias for a single sketch can be expanded as follows:
\begin{align*}
    \| \Exs [A\hat{x}_k]-Ax^* \|_2 &= \|\Exs[A(A^TS_k^TS_kA)^{-1}A^TS_k^TS_kb] - Ax^* \|_2 \\
    &= \|\Exs[A(A^TS_k^TS_kA)^{-1}A^TS_k^TS_k(Ax^*+b^\perp)] - Ax^* \|_2 \\
    &= \|\Exs[A(A^TS_k^TS_kA)^{-1}A^TS_k^TS_k b^\perp ] \|_2 \\
    &= \|U \Exs[(U^TS_k^TS_kU)^{-1}U^TS_k^TS_kb^\perp \|_2 \\
    &= \| \Exs[(U^TS_k^TS_kU)^{-1}U^TS_k^TS_kb^\perp \|_2 \\
    &= \|\Exs[Qz] \|_2,
\end{align*}
where we define $Q \defn (U^TS_k^TS_kU)^{-1} $ and $z \defn U^TS_k^TS_kb^\perp$. The term $\|\Exs[Qz] \|_2^2$ can be upper bounded as follows:
\begin{align*}
    \|\Exs[Qz] \|_2^2 &= \Exs[Qz]^T \Exs[Qz] \\
    &= \Exs_{S_k}[Qz]^T \Exs_{S_k^\prime}[Q^\prime z^\prime] \\
    &= \Exs_{S_k} \Exs_{S_k^\prime} [z^T Q Q^\prime z^\prime ] \\
    &= \frac{1}{2} \Exs_{S_k} \Exs_{S_k^\prime} [(z+z^\prime)^TQQ^\prime (z+z^\prime) - z^TQQ^\prime z - {z^\prime}^T QQ^\prime z^\prime] \\
    &\leq \frac{1}{2} \Exs_{S_k} \Exs_{S_k^\prime} [ \left(\|z+z^\prime\|_2^2 (1+\epsilon)^2 - \|z\|_2^2(1-\epsilon)^2 - \|z^\prime \|_2^2(1-\epsilon)^2 \right) ] \\
    &=  \Exs_{S_k} \Exs_{S_k^\prime} [ \left( \|z\|_2^2 2\epsilon + \|z^\prime\|_2^2 2\epsilon + z^Tz^\prime (1+\epsilon)^2 \right)] \\
    &= 2\epsilon \Exs_{S_k} \Exs_{S_k^\prime} [\|z\|_2^2 ]  +  2\epsilon \Exs_{S_k} \Exs_{S_k^\prime} [\|z^\prime \|_2^2]  +  (1+\epsilon)^2 \Exs_{S_k} \Exs_{S_k^\prime} [z^Tz^\prime ] \\
    &= 4\epsilon \Exs [\|z\|_2^2 ]  +  (1+\epsilon)^2 \|\Exs [z ]\|_2^2,
 \end{align*}
where the inequality in the fifth line follows from the assumption $(1-\epsilon)I_d \preceq Q \preceq (1+\epsilon)I_d$ and some simple bounds for the minimum and maximum eigenvalues of the product of two positive definite matrices. Furthermore, the expectation of $z$ is equal to zero because $\Exs[z] = \Exs[U^TS_k^TS_kb^\perp]=U^T\Exs[S_k^TS_k]b^\perp=U^Tb^\perp=0$. Hence we have 
\begin{align*}
    \| \Exs [A\hat{x}_k]-Ax^* \|_2 \leq \sqrt{4\epsilon \Exs[\|z\|_2^2]}.
\end{align*}

% By Lemma \ref{expectation_upper_bound} and Lemma \ref{bound_z_times_indicator}, we have:
% \begin{align*}
%     \|\Exs[Qz] \|_2^2 &\leq 4\epsilon \mprob[E] \Exs_S [\|z\|_2^2]  +  (1+\epsilon)^2 \mprob[E]^2 \Exs [\|z\|_2^2] \\
%     &= (4\epsilon \mprob[E] + (1+\epsilon)^2 \mprob[E]^2) \Exs_S [\|z\|_2^2] \\
%     %&\leq (4\epsilon + (1+\epsilon)^2) \Exs_S [\|z\|_2^2] \\
%     &\leq (1+7\epsilon) \Exs_S [\|z\|_2^2]
% \end{align*}
% Finally, we obtain the upper bound on the norm of the bias as follows:
% \begin{align*}
%     \| \Exs [A\tilde{x}]-Ax^* \|_2 &\leq  \sqrt{(1+7\epsilon) \Exs_S [\|z\|_2^2]} + \|Ax^*\|_2 \mprob[E^c]
% \end{align*}
\end{proof}

\proof[\textbf{Proof of Lemma \ref{bound_z_norm_ROS}}]
For the randomized Hadamard sketch (ROS), the term $\Exs[ \| z\|_2^2]$ can be expanded as follows:
\begin{align*}
\Exs [\|z\|_2^2] &= \Exs \left[ {b^\perp}^T \frac{1}{m}\sum_{i=1}^m s_i s_i^T UU^T \frac{1}{m}\sum_{j=1}^m s_j s_j^T b^{\perp} \right] \\
&= \Exs \left[ \frac{1}{m^2} \sum_{i=1}^m \sum_{j=1}^m {b^\perp}^T s_i s_i^T UU^T s_j s_j^T b^{\perp} \right]\\
&=  \frac{1}{m^2} \sum_{1\le i=j\le m} {b^\perp}^T \Exs \left[ s_i s_i^T UU^T s_j s_j^T\right] b^{\perp} + \frac{1}{m^2} \sum_{i\neq j,\,1\le i,j\le m}  {b^\perp}^T \Exs [s_i s_i^T] UU^T \Exs [s_j s_j^T] b^{\perp}\\
&= \frac{m}{m^2} \, {b^\perp}^T \Exs \left[ s_1 s_1^T UU^T s_1 s_1^T \right] b^{\perp} + \frac{1}{m^2} \sum_{i\neq j,\,1\le i,j\le m}  {b^\perp}^T I_n UU^T I_n b^{\perp} \\
&= \frac{1}{m} \, {b^\perp}^T \Exs \left[ s_1 s_1^T UU^T s_1 s_1^T \right] b^{\perp},
\end{align*}
where we have used the independence of $s_i$ and $s_j$, $i \neq j$ in going from the second line to the third line. This is true because of the assumption that the matrix $P$ corresponds to sampling with replacement.

\begin{align*}
    {b^\perp}^T \Exs \left[ s_1 s_1^T UU^T s_1 s_1^T \right] b^{\perp} &= \Exs [ (s_1^TUU^Ts_1) (s_1^Tb^\perp {b^\perp}^Ts_1) ] \\
    &= \Exs [ (s_1^TUU^Ts_1) ({b^\perp}^Ts_1)^2 ] \\
    &= \frac{1}{n} \sum_{i=1}^n \Exs[(h_i^TDUU^TDh_i)({b^\perp}^TDh_i)^2],
\end{align*}
where the row vector $h_i^T$ corresponds to the $i$'th row of the Hadamard matrix $H$. We also note that the expectation in the last line is with respect to the randomness of $D$.

Let us define $r$ to be the column vector containing the diagonal entries of the diagonal matrix $D$, that is, $r \coloneqq [D_{11}, D_{22}, ..., D_{nn}]^T$. Then, the vector $Dh_i$ is equivalent to $Diag(h_i)r$ where $Diag(h_i)$ is the diagonal matrix with the entries of $h_i$ on its diagonal.
\begin{align*}
    \frac{1}{n} \sum_{i=1}^n \Exs[(h_i^TDUU^TDh_i)({b^\perp}^TDh_i)^2] &= \frac{1}{n} \sum_{i=1}^n \Exs[(r^TDiag(h_i)UU^TDiag(h_i)r)({b^\perp}^TDiag(h_i)r)^2] \\
    &= \frac{1}{n} \sum_{i=1}^n \Exs[{b^\perp}^T Diag(h_i) r (r^TPr)  r^T Diag(h_i) b^\perp] \\
    &= \frac{1}{n} \sum_{i=1}^n {b^\perp}^T Diag(h_i) \Exs[ r (r^TPr) r^T ]Diag(h_i) b^\perp ,
\end{align*}
where we have defined $P \coloneqq Diag(h_i) UU^T Diag(h_i)$. It follows that $\Exs[ r (r^TPr) r^T ] = 2P - 2 Diag(P) + \tr( P) I_n$. Here, $Diag(P)$ is used to refer to the diagonal matrix with the diagonal entries of $P$ as its diagonal. 

The trace of $P$ can be easily computed using the cyclic property of matrix trace as $\tr (P) = \tr (Diag(h_i) UU^T Diag(h_i)) = \tr(U^T Diag(h_i) Diag(h_i) U) = \tr(U^TU) = \tr(I_d) = d$.

% -------
% \begin{align*}
% \mbox{\red{ (*) this step can be improved}}
% &\le\frac{1}{m} \, {b^\perp}^T \frac{1}{n} \sum_{i=1}^n s_i  s_i^T b^{\perp} \max_{i=1,...,n} (s_i^T UU^T s_i) \\
% &=\frac{1}{m} \, {b^\perp}^T b^{\perp} \max_{i=1,...,n} (s_i^T UU^T s_i) \\
% &=\frac{1}{m} \| {b^\perp}\|_2^2 \max_{i=1,...n} \|U^TDH p_i\|_2^2
% \end{align*}
% %
% \mbox{\red{(*)}}'s expectation can be computed as
% %
% \begin{align}
%     \sum_i \Exs ({b^\perp}' D h_i)^2 (h_i^T D  UU^T D h_i)  &= \sum_i \Exs   ({b^\perp}' diag(h_i) r)^2 (r diag(h_i)  UU^T diag(h_i) r \\
%     &= \sum_i \Exs \sum_{ijkl} r_i r_j r_k r_l a_i a_j q_{ij}\\
%     & = \sum_i \Exs \sum_{j} a_j a_j q_{jj} + \Exs \sum_{kl} a_k a_k q_{ll} + ....
% \end{align}
% %
% %
% where $r = vec(D)$, $a_i = b^\perp diag(h)$, $q_ij=diag(h_i) UU^T diag(h_i)$.\\
% %
% Another way to bound the term in \mbox{\red{(*)}}: Let $r = vec(D)$ i.i.d. $\pm 1$. $\Exs r r^T r^T Q r = I \trace Q  + Q + Q^T - 2diag(Q)$ for any matrix Q. We set $Q=Diag(h_i) UU^T Diag(h_i)$, which satisfies $\trace Q = \trace U^T Diag(h_i)Diag(h_i) U = \trace U^T U = \trace I_d = d$ using the cyclic property of matrix trace.

We note that the term $Diag(P)$ can be simplified as $Diag(P)_{jj}=\|\tilde{u}_j \|_2^2$ where $\tilde{u}_j^T$ is the $j$'th row of $U$. This leads to
\begin{align*}
    {b^\perp}^T diag(P) b^\perp &= \sum_{j=1}^n (b_j^\perp)^2 \|\tilde{u}_j \|_2^2 \\
    &\geq \sum_{j=1}^n (b_j^\perp)^2 \min_i \|\tilde{u}_i \|_2^2 \\
    &= \|b^\perp \|_2^2 \min_i \|\tilde{u}_i \|_2^2.
\end{align*}

Going back to $\Exs [\|z \|_2^2]$,
\begin{align*}
    \Exs [\|z \|_2^2] &= \frac{1}{mn} {b^\perp}^T  n Diag(h_i) \left(2P-2Diag(P) + \tr(P)I_n\right)Diag(h_i) b^\perp \\
    &= \frac{d}{m} \|b^\perp \|_2^2 - \frac{2}{m} {b^\perp}^T Diag(P) b^\perp \\
    &\leq \frac{d}{m} \|b^\perp \|_2^2 - \frac{2}{m} \|b^\perp \|_2^2 \min_i \|\tilde{u}_i \|_2^2 \\
    &= \frac{1}{m} \|b^\perp \|_2^2 (d-2\min_i \|\tilde{u}_i \|_2^2) \\
    &= \frac{d}{m} \left(1-\frac{2\min_i \|\tilde{u}_i \|_2^2}{d} \right) f(x^*).
\end{align*}
%
%where $m=$ \todo{what exactly do we set m to?}
%\todo{we can set $m=d \log(2n)/\epsilon)$}.
%
% Next taking expectations over the sign variables $D$ we obtain
% %
% \begin{align*}
% \Exs_D \Exs \|z\|_2^2 &\le \frac{1}{m} \| {b^\perp}\|_2^2 \rank(A) \log(2n)\\
% &\le \epsilon f(x^*).
% \end{align*}
% %
% The above inequality follows from the fact that $\sum_{j} H_{ij}D_{jj} U_{j}$ is sub-Gaussian with parameter $\|U\|_F^2 = \trace UU^T = rank(A)$.
\qed

\proof[\textbf{Proof of Lemma \ref{bound_z_norm_uniform}}]
Note that $\Exs[S_k^TS_k] = \frac{1}{m} \sum_{i=1}^m s_i s_i^T = I_n$ where $s_i \in \mathbb{R}^n$ is the column vector corresponding to the $i$'th row of $S_k$ scaled by $1/\sqrt{m}$. Note that because of this particular scaling, $\Exs[s_is_i^T] = I_n$ holds.
For uniform sampling with replacement, we have
\begin{align*}
    \Exs [\|z\|_2^2] &= \frac{1}{m^2}\, \Exs\left[ {b^\perp}^T \sum_{i=1}^m s_i s_i^T UU^T \sum_{j=1}^m s_j s_j^T b^{\perp} \right] \\
    &= \frac{1}{m^2}\, \Exs\left[ \sum_{i=1}^m \sum_{j=1}^m {b^\perp}^T s_i s_i^T UU^T s_j s_j^T b^{\perp} \right] \\
    &= \frac{1}{m^2}\, \sum_{1\le i=j\le m} {b^\perp}^T \Exs [s_i s_i^T UU^T s_j s_j^T] b^{\perp} + \frac{1}{m^2}\,\sum_{i\neq j,\,1\le i,j\le m}  {b^\perp}^T \Exs [s_i s_i^T] UU^T \Exs [s_j s_j^T] b^{\perp}\\
    &= \frac{1}{m}\, {b^\perp}^T \Exs [s_1 s_1^T UU^T s_1 s_1^T] b^{\perp} + \frac{1}{m^2}\, \sum_{i\neq j,\,1\le i,j\le m}  {b^\perp}^T I_n UU^T I_n b^{\perp} \\
    &= \frac{1}{m}\, {b^\perp}^T \Exs [s_1 s_1^T UU^T s_1 s_1^T] b^{\perp} \\
    &= \frac{1}{m}\, {b^\perp}^T n^2 \frac{1}{n} \sum_{i=1}^n e_ie_i^TUU^Te_ie_i^T b^\perp \\
    &= \frac{n}{m}\, {b^\perp}^T \sum_{i=1}^n \|\tilde{u}_i\|_2^2 e_i e_i^T b^\perp \\
    &= \frac{n}{m}\, {b^\perp}^T Diag( \|\tilde{u}_i \|_2^2 ) b^\perp \\
    &= \frac{n}{m}\, \sum_{i=1}^n {b_i^\perp}^2 \|\tilde{u}_i \|_2^2 \\
    &\leq \frac{n}{m}\, \sum_{i=1}^n {b_i^\perp}^2 \max_j \|\tilde{u}_j \|_2^2 \\
    &= \frac{n}{m}\, f(x^*) \max_i \|\tilde{u}_i \|_2^2.
\end{align*}
For uniform sampling without replacement, the rows $s_i$ and $s_j$ are not independent which can be seen by noting that given $s_i$, we know that $s_j$ will have its nonzero entry at a different place than $s_i$. Hence, differently from uniform sampling with replacement, the following term will not be zero:
\begin{align*}
    \frac{1}{m^2}\,\sum_{i\neq j,\,1\le i,j\le m}  {b^\perp}^T \Exs [s_i s_i^T UU^T s_j s_j^T] b^{\perp} &= \frac{m^2-m}{m^2} {b^\perp}^T \Exs[s_1s_1^TUU^Ts_2s_2^T] b^\perp \\
    &= \frac{m-1}{m} {b^\perp}^T n^2 \frac{1}{n^2-n} \sum_{i\neq j, 1 \leq i,j \leq n} e_ie_i^TUU^Te_je_j^Tb^\perp \\
    &= \frac{m-1}{m} \frac{n}{n-1} {b^\perp}^T \sum_{i\neq j, 1 \leq i,j \leq n} e_i\tilde{u}_i^T \tilde{u}_j e_j^Tb^\perp \\
    &= \frac{m-1}{m} \frac{n}{n-1} {b^\perp}^T (UU^T - Diag(\|\tilde{u}_i \|_2^2) b^\perp \\
    &= \frac{m-1}{m} \frac{n}{n-1} (0 - \sum_{i=1}^n {b_i^\perp}^2 \|\tilde{u}_i \|_2^2) \\
    &= -\frac{m-1}{m} \frac{n}{n-1}\sum_{i=1}^n {b_i^\perp}^2 \|\tilde{u}_i \|_2^2.
\end{align*}
It follows that for uniform sampling without replacement, we obtain
\begin{align*}
    \Exs[\|z\|_2^2] &= \left (\frac{n}{m} - \frac{m-1}{m} \frac{n}{n-1} \right) \sum_{i=1}^n {b_i^\perp}^2 \|\tilde{u}_i \|_2^2 \\
    &= \frac{n}{m} \frac{n-m}{n-1} \sum_{i=1}^n {b_i^\perp}^2 \|\tilde{u}_i \|_2^2 \\
    &\leq \frac{n}{m} \frac{n-m}{n-1}\, f(x^*) \max_i \|\tilde{u}_i \|_2^2.
\end{align*}
\qed

\proof[\textbf{Proof of Lemma \ref{bound_z_norm_leverage}}]
We consider leverage score sampling with replacement. The rows $s_i$, $s_j$ $i \neq j$ are independent because sampling is with replacement. For leverage score sampling, the term $\Exs [\|z\|_2^2]$ is upper bounded as follows:
\begin{align*}
    \Exs [\|z\|_2^2] &= \frac{1}{m^2}\, \Exs\left[ {b^\perp}^T \sum_{i=1}^m s_i s_i^T UU^T \sum_{j=1}^m s_j s_j^T b^{\perp} \right] \\
    &= \frac{1}{m^2}\, \Exs\left[ \sum_{i=1}^m \sum_{j=1}^m {b^\perp}^T s_i s_i^T UU^T s_j s_j^T b^{\perp} \right] \\
    &= \frac{1}{m^2}\, \sum_{1\le i=j\le m} {b^\perp}^T \Exs [s_i s_i^T UU^T s_j s_j^T] b^{\perp} + \frac{1}{m^2}\, \sum_{i\neq j,\,1\le i,j\le m}  {b^\perp}^T \Exs [s_i s_i^T] UU^T \Exs [s_j s_j^T] b^{\perp} \\
    &= \frac{1}{m}\, {b^\perp}^T \Exs [s_1 s_1^T UU^T s_1 s_1^T] b^{\perp} + \frac{m^2-m}{m^2} {b^\perp}^T \Exs [s_1 s_1^T] UU^T \Exs [s_1 s_1^T] b^{\perp} \\
    &= \frac{1}{m}\, {b^\perp}^T \sum_{i=1}^n \frac{\ell_i}{d} \frac{d}{\ell_i}e_ie_i^TUU^T \frac{d}{\ell_i}e_ie_i^T b^\perp  +  \frac{m^2-m}{m^2} {b^\perp}^T I_n UU^T I_n b^{\perp}\\
    &= \frac{1}{m}\, {b^\perp}^T \sum_{i=1}^n \frac{d}{\ell_i} \ell_i e_ie_i^T b^\perp  \\
    &= \frac{d}{m}\, {b^\perp}^T \sum_{i=1}^n e_ie_i^T b^\perp  \\
    &= \frac{d}{m}\, \|b^\perp \|_2^2 \\
    &= \frac{d}{m} f(x^*).
\end{align*}
\qed
\subsection{Differential Privacy}
The proof of the privacy result mainly follows due to Theorem \ref{thm:diff_priv_cited} (stated below for completeness) which is a result from \cite{sheffet15privacy}.

\proof[\textbf{Proof of Theorem \ref{thm:diff_priv_result}}]
For some $\varepsilon$, $\delta$ and matrix $A_c$, if there exist values for $m$ such that the smallest singular value of $A_c$ satisfies $\sigma_{\min}(A_c) \geq w$, then using Theorem \ref{thm:diff_priv_cited}, we find that the sketch size $m$ has to satisfy the following for $(\varepsilon, \delta)$-differential privacy: 
\begin{align} \label{eq:sketch_size_privacy}
    m &\leq  \frac{1}{8\ln(4/\delta)} \left(\left(\frac{\sigma_{\min}^2}{B^2} - 1\right) \frac{1}{\frac{1}{\varepsilon}+\frac{1}{\ln(4/\delta)}} - 2\ln(4/\delta) \right)^2 \nonumber \\
    &= \frac{1}{8\beta} \left(\left(\frac{\sigma_{\min}^2}{B^2} - 1\right) \frac{\varepsilon \beta}{\varepsilon + \beta} - 2\beta \right)^2,
\end{align}
where we have set $\delta = 4/e^\beta$ in the second line. For the first line to follow from Theorem \ref{thm:diff_priv_cited}, we also need the condition $\frac{\sigma_{\min}^2}{B^2} \geq 3 + 2\frac{\beta}{\varepsilon}$ to be satisfied. We now substitute $B=B_0 \sqrt{d}$ and $\sigma_{\min} = \sigma_0 \sqrt{n}$ to obtain the simplified condition
\begin{align*}
    \frac{n}{d} \geq \left(3+2 \frac{\beta}{\varepsilon} \right) \frac{B_0^2}{\sigma_0^2},
\end{align*}
where $B_0$ and $\sigma_0$ are constants. Assuming this condition is satisfied, then we pick the sketch size $m$ as \eqref{eq:sketch_size_privacy} which can also be simplified:
\begin{align*}
    m = O\left(\beta \frac{n^2}{d^2} \frac{\varepsilon^2}{(\varepsilon + \beta)^2} \right).
\end{align*}
Note that the above arguments are for the privacy of a single sketch (i.e., $S_kA_c$). In the distributed setting where the adversary can attack all of the sketched data $S_1A_c, ..., S_qA_c$, we can consider all of the sketched data to be a single sketch with size $mq$. Based on this argument, we can pick the sketch size as
\begin{align*}
    m = O\left(\frac{\beta}{q} \frac{n^2}{d^2} \frac{\varepsilon^2}{(\varepsilon + \beta)^2} \right).
\end{align*}

% we should have d+1 instead of d
%%%%%%

% If we fix $\delta = 4/e^\beta$, then $\ln(4/\delta) = \beta$. Then, $m$ can be picked:
% \begin{align*}
%     m &\leq \frac{1}{q} \frac{1}{24} \left(\left(\frac{\sigma_{\min}^2}{B^2} - 1\right) \frac{3\epsilon}{3+\epsilon} - 6 \right)^2 \\
%     &= O\left(\frac{\epsilon^2}{q} \right)
% \end{align*}

% Picking $m = O\left(\frac{\epsilon^2}{q} \right)$ and assuming $d << m$, we obtain
% \begin{align*}
%     \frac{\Exs[f(\bar{x})]-f(x^*)}{f(x^*)} &= \frac{1}{q} \frac{d}{m-d-1} \\
%     &= O\left(\frac{1}{\epsilon^2}\right)
% \end{align*}
\qed

\btheos[Differential privacy for random projections \cite{sheffet15privacy}] \label{thm:diff_priv_cited}
Fix $\epsilon > 0$ and $\delta \in (0, 1/e)$. Fix $B > 0$. Fix a positive integer $m$ and let $w$ be such that
\begin{align} \label{eq:def_w}
    w^2 = B^2 \left( 1 + \frac{1+\frac{\epsilon}{\ln(4/\delta)}}{\epsilon} \left( 2 \sqrt{2m \ln(4/\delta)} + 2\ln(4/\delta) \right)\right).
\end{align}
Let $A$ be an $(n\times d)$-matrix with $d < m$ and where each row of $A$ has bounded $\ell_2$-norm of $B$. Given that $\sigma_{\min}(A) \geq w$, the algorithm that picks an $(m \times n)$-matrix $R$ whose entries are iid samples from the normal distribution $\mathcal{N}(0,1)$ and publishes the projection $RA$ is $(\epsilon, \delta)$-differentially private.
\etheos

We note that Remark 1 is a direct consequence of of Theorem \ref{thm:diff_priv_result} and Theorem \ref{Thm:AvgGaussian}.

%%%%%%%
\subsection{Proofs for Right Sketch}
\proof[\textbf{Proof of Lemma \ref{gaussian_one_sketch_rightsketch}}]
It follows that conditioned on $AS_k^T$, we have 
\begin{align*}
    \hat x_k \sim \mathcal{N}\Big(x^*, P_{\mbox{Null}(A)}\| AS_k^T (AS_k^TS_kA^T)^{-1} b \|_2^2 \Big)\,.
\end{align*}
Noting that $\Exs [(AS_k^TS_kA^T)^{-1}] = AA^T \frac{m}{m-n-1} $, taking the expectation and noting that $\mbox{tr} P_{\mbox{Null}(A)}=d-n$, we obtain
\begin{align}
    \Exs [\| \hat x_k - x^* \|_2^2] &= \frac{d-n}{m-n-1} b^T (AA^T)^{-1} b = \frac{d-n}{m-n-1} f(x^*). 
\end{align} 
\qed

\textbf{Error for the averaged solution:} Note that Lemma \ref{gaussian_one_sketch_rightsketch} establishes the expected error for the single sketch estimator $\hat{x}_k$. We now present an exact formula for the averaged solution $\bar{x}$. We first note the unbiasedness of the estimates, i.e., $\Exs[\hat{x}_k] = x^*$. Next, following similar steps to those used in proving Lemma \ref{expected_obj_val_diff} along with the unbiasedness of the estimates, we obtain
\begin{align*}
    \Exs[\|\bar{x} - x^* \|_2^2] &= \frac{1}{q} \Exs[\|\hat{x}_k - x^* \|_2^2] \\
    &= \frac{1}{q} \frac{d-n}{m-n-1} f(x^*).
\end{align*}
Hence, for the Gaussian right sketch, we establish the approximation error as
\begin{align}
    \frac{\Exs[f(\bar{x})] - f(x^*)}{f(x^*)} = \frac{1}{q} \frac{d-n}{m-n-1}.
\end{align}
\end{document}